\documentclass[english,eqsecnum,nofootinbib,superscriptaddress,preprintnumbers]{revtex4-2}
\usepackage[T1]{fontenc}
\usepackage[utf8]{inputenc}
\usepackage{color}
\usepackage{colortbl}
\usepackage{babel}
\usepackage{verbatim}
\usepackage{float}
\usepackage{mathtools}
\usepackage{bm}
\usepackage{amsmath}
\usepackage{empheq}
\usepackage[pdfusetitle,
 bookmarks=true,bookmarksnumbered=false,bookmarksopen=false,
 breaklinks=false,pdfborder={0 0 1},backref=false,colorlinks=true]
 {hyperref}

\makeatletter

\providecommand{\tabularnewline}{\\}

\usepackage{colortbl}
\usepackage{xcolor}
\usepackage[normalem]{ulem}
\usepackage{babel}
\usepackage{bm}
\usepackage{amssymb}

\definecolor{XGDarkRed}{RGB}{200,0,0}

\definecolor{deletegray}{RGB}{160,160,160}

\providecommand{\tabularnewline}{\\}

\makeatother

\begin{document}
\title{Bi-scalar-tensor theory: a spatially covariant gravity approach}
\author{Jia-Jun Chen}
\affiliation{School of Physics, Sun Yat-sen University, Guangzhou 510275, China}

\author{Xian Gao}
\email[Email: ]{gaoxian@mail.sysu.edu.cn}
\affiliation{School of Physics, Sun Yat-sen University, Guangzhou 510275, China}
\affiliation{Guangdong Provincial Key Laboratory of Quantum Metrology and Sensing,\\ Sun Yat-sen University, Zhuhai 519082, China}

\author{Tsutomu Kobayashi}
\email[Email: ]{tsutomu@rikkyo.ac.jp}
\affiliation{Department of Physics, Rikkyo University, Toshima, Tokyo 171-8501, Japan}

\date{September 20, 2026}
\begin{abstract}
We propose a novel method for constructing ghost-free multi-field higher-derivative scalar-tensor theories. The idea is to start from spatially covariant gravity and couple it to an additional scalar field. The resulting spatially covariant scalar field theory can be understood as a bi-scalar-tensor theory in the unitary gauge of one scalar. Within a broad class in which metric velocities enter through the extrinsic curvature and the additional scalar enters through temporal derivatives up to second order, we perform a Hamiltonian constraint analysis. We show that the theory generally propagates five physical degrees of freedom (DOFs), i.e., two tensor modes and three scalar modes. One of the scalar modes is the unwanted mode associated with the additional scalar. We then derive a degeneracy condition and an additional consistency condition. When both conditions hold, the unwanted mode is eliminated and the theory propagates four degrees of freedom. The resulting theories fall into two cases, depending on whether the preservation of the tertiary constraint closes without a new constraint or generates a quaternary constraint. We also study the two-field disformal transformation, derive its invertibility condition, and express it in the unitary gauge. Our construction provides a systematic approach to constructing bi- and multi-scalar-tensor theories with higher derivatives.
\end{abstract}

\preprint{RUP-26-20}

\maketitle

\section{Introduction}

The accelerated expansion of the Universe, the phenomenology of the early Universe, and the possibility of testing gravity across cosmological and strong-field scales continue to motivate extensions of general relativity (GR) \citep{Koyama:2015vza,Ferreira:2019xrr,Arai:2022ilw}. Scalar-tensor theories provide one of the simplest settings in which the gravitational sector can acquire an additional dynamical degree of freedom. Their construction becomes delicate once higher derivatives are admitted, because a nondegenerate higher-time-derivative Lagrangian generically carries an Ostrogradsky instability \citep{Woodard:2015zca}. Requiring second-order field equations led from $k$-inflation and the Galileon to the covariant and generalized Galileons \citep{Armendariz-Picon:1999hyi,Nicolis:2008in,Deffayet:2009wt,Deffayet:2009mn,Deffayet:2011gz}, which are equivalent to Horndeski's most general four-dimensional scalar-tensor theory with second-order equations of motion \citep{Horndeski:1974wa,Kobayashi:2011nu}. The second-order requirement is sufficient but not necessary. Beyond-Horndeski and degenerate higher-order scalar-tensor (DHOST) theories use a degenerate kinetic structure to retain the desired number of degrees of freedom despite higher-order equations \citep{Gleyzes:2014dya,Gleyzes:2014qga,Langlois:2015cwa,Motohashi:2016ftl} (see \citep{Kobayashi:2019hrl} for a review). Recent works have extended the construction of healthy higher-derivative scalar-tensor interactions beyond the conventional DHOST framework \citep{Takahashi:2021ttd,Takahashi:2022mew,Michiwaki:2026xru,Gorji:2026nes}.

The extension of these constructions to multiple scalar fields remains less understood. Bi- and multi-Galileons were constructed in flat spacetime, including theories motivated by higher-codimension branes and internal symmetries \citep{Deffayet:2010zh,Padilla:2010de,Padilla:2010tj,Padilla:2010ir,Hinterbichler:2010xn,Zhou:2010di}. Their covariant extensions and multi-field versions of generalized $G$-inflation revealed interactions that are absent in a direct replication of the single-field theory \citep{Padilla:2012dx,Kobayashi:2013ina,Akama:2017jsa}. Further classifications have exposed additional antisymmetric or multi-flavor structures \citep{Allys:2016hfl,Bogers:2018kuw,Roest:2019oiw,Kampf:2020tne,Aoki:2021kla}. 
The general second-order field equations for two scalar fields were derived in \cite{Ohashi:2015fma}.  Horndeski recently revisited the construction and emphasized the remaining gap between known field equations and a complete general action \citep{Horndeski:2024hee}. A recent proposal based on closure under invertible disformal transformations offers another route toward multi-Horndeski actions, but its complete equivalence to the general bi-scalar field equations and its extension to more fields remain open \citep{Katayama:2025hnd}. Beyond the second-order class, the Hamiltonian analysis of genuinely multi-field higher-derivative theories involves field-space degeneracy and additional consistency and rank conditions with no direct single-field analogue \citep{BouzariNezhad:2026zsj}. However, a general constructive classification of healthy higher-derivative multi-scalar-tensor theories is still lacking.

Several approaches have been developed to eliminate unwanted degrees of freedom. One may restrict the equations of motion to second order, or instead impose degeneracy of the kinetic matrix and follow the full constraint chain, as in DHOST theories. Geometric probe-brane constructions generate Galileon and DBI-Galileon interactions from the induced geometry and Lovelock invariants or their boundary terms \citep{Lovelock:1971yv,Myers:1987yn,Miskovic:2007mg,deRham:2010eu,Goon:2011uw}. Higher-codimension embeddings naturally produce several scalars and curved-background generalizations \citep{Hinterbichler:2010xn,Goon:2011qf,Goon:2011xf,Burrage:2011bt}. Invertible field redefinitions provide another constructive tool. Starting from the disformal relation introduced in Ref.~\cite{Bekenstein:1992pj}, transformations of Horndeski and higher-order scalar-tensor theories have been used to organize apparently different representations while preserving the number of degrees of freedom when the map is regular and invertible \citep{Bettoni:2013diz,Zumalacarregui:2013pma,Deffayet:2020ypa}. Multi-field extensions have been studied both for nonlinear cosmological perturbations and for two-field dynamics \citep{Watanabe:2015uqa,Firouzjahi:2018xob}. Recent work also shows explicitly that consistency of a two-scalar disformal coupling can impose a degenerate field-space structure \citep{Domenech:2025gao}. None of these methods makes a completely general higher-derivative multi-field action automatically healthy. The relevant second-order, degeneracy, invertibility, and constraint-consistency requirements must be checked in the class under consideration.

Spatially covariant gravity (SCG) provides a different and efficient formulation of the single-field problem \citep{Gao:2014soa,Gao:2014fra}. If a scalar field $\phi$ has a timelike gradient, one may choose its unitary gauge, $\phi=\phi(t)$, so that the scalar fluctuation is encoded in the metric and the action is built from geometric quantities on the constant-$\phi$ hypersurfaces. In this description temporal diffeomorphisms, or refoliation invariance, are not manifest, whereas spatial diffeomorphisms are preserved. The additional scalar mode can therefore be viewed as arising from the breaking of time reparametrization invariance. This logic is shared by the effective field theory of inflation and dark energy \citep{Creminelli:2006xe,Cheung:2007st,Gubitosi:2012hu}, and is closely related to the preferred-foliation structure of Ho\v{r}ava gravity and its healthy extensions \citep{Horava:2009uw,Blas:2009qj,Blas:2009yd,Blas:2010hb}. Similar symmetry-breaking structures also appear in effective theories of gravity, including ghost condensation \citep{Arkani-Hamed:2003pdi}, nonlinear massive gravity \citep{Boulware:1972yco,deRham:2010kj,Hassan:2011hr}, and solid inflation \citep{Endlich:2012pz}. The utility of SCG is not that every spatially covariant action is ghost free, but that the separation of Lie and spatial derivatives makes the kinetic structure and the conditions eliminating unwanted modes more transparent.

The original SCG construction has been extended in several directions. Disformal and canonical transformations were analyzed in Refs.~\cite{Fujita:2015ymn,Saitou:2016lvb}, while actions involving the velocity of the lapse require additional degeneracy and consistency conditions \citep{Gao:2018znj,Gao:2019lpz,Lin:2020nro}. Auxiliary or nondynamical scalar fields provide further ways of changing the constraint structure \citep{Gao:2018izs,Wang:2024hfd,Zhu:2025bcu}. Restoring temporal diffeomorphism invariance by the Stueckelberg procedure maps an SCG action to a generally covariant scalar-tensor theory. The resulting covariant correspondence, including classifications of higher-derivative monomials and the relation between unitary-gauge and covariant degeneracy conditions, has been developed in Refs.~\cite{Gao:2020juc,Gao:2020qxy,Gao:2020yzr,Hu:2021bbo,Hu:2024hzo}. This correspondence permits higher spatial derivatives in unitary gauge without thereby introducing higher time derivatives. The resulting covariant theory is healthy if the corresponding constraint conditions are satisfied and the unitary-gauge description is valid.
 
SCG has also been applied to cosmological perturbations and gravitational-wave propagation. The features of Lorentz and parity violation were studied in Refs.~\cite{Gao:2019liu,Zhu:2022dfq,Zhu:2022uoq,Zhu:2023rrx}, and scalar-induced gravitational waves have recently been calculated in the SCG framework \citep{Jiang:2025ysb}. A particularly active branch seeks theories with only the two tensorial degrees of freedom. Hamiltonian, perturbative, and auxiliary-constraint constructions have produced general conditions and explicit candidates \citep{Gao:2019twq,Hu:2021yaq,Yao:2020tur,Yao:2023qjd,Wang:2024hfd,Zhu:2025bcu,Yu:2026fgn}. Their cosmological properties, matter couplings, and observational consequences have been investigated in Refs.~\cite{Iyonaga:2021yfv,Hiramatsu:2022ahs,Chakraborty:2023jek}. Black-hole perturbations provide a complementary strong-field application \citep{Saito:2023bhn}. The geometric framework has furthermore been generalized to nonmetricity \citep{Yu:2024drx}, to parity-violating operators at increasingly high derivative order \citep{Hu:2024hzo,Song:2026ajt}, and to spatially covariant vector-field theories \citep{Li:2025qbv}. These examples illustrate the flexibility of the framework of spatially covariant gravity, while also showing that the number of propagating degrees of freedom is determined by the constraint structure of the specific SCG theory.

A previous extension of SCG to multiple scalar fields introduced a separate foliation for each field and coupled the corresponding hypersurface geometries \citep{Yu:2024sed}. The multi-hypersurface construction treats the scalar fields symmetrically and makes genuinely multi-field geometric interactions manifest. For a specific bi-scalar model, a linear cosmological perturbation analysis found two tensor and two scalar modes. That result, however, establishes the mode content only around the background and at the perturbative order analyzed. The general interacting-hypersurface action has not yet been subjected to a complete nonlinear Hamiltonian constraint analysis, so whether the full construction eliminates unwanted modes remains an open question. 

In this work we propose a complementary, asymmetric construction. We begin with a single-foliation SCG theory associated with a scalar $\phi$ and couple an additional scalar field $\psi$ directly to the spatially covariant geometric variables. From the covariant viewpoint, this is a bi-scalar-tensor theory written in the unitary gauge $\phi=t$. From the gauge-fixed viewpoint, it is SCG coupled to a scalar field, which we call a spatially covariant scalar field theory. The two descriptions distinguish the general action from its healthy subclasses. The broad spatially covariant action is  the starting point for the constraint analysis, which is not assumed to be ghost free by itself. This setup thus differs from treating $\psi$ as an ordinary minimally coupled matter field, for which preservation of the gravity-sector degeneracy is nontrivial \citep{Deffayet:2020ypa,Takahashi:2022ctx}. This setup is also different from effective descriptions in which one scalar is assigned to dark energy and the other to a matter sector \citep{Gergely:2014rna}.

In the single-foliation formulation, Lie derivatives along the preferred normal are separated from spatial derivatives, allowing higher spatial derivatives to be included while keeping explicit control over the highest time derivatives. The usual Arnowitt-Deser-Misner variables are also the natural variables for a nonlinear Hamiltonian analysis. The price is that the two scalar fields are not treated on an equal footing. The scalar field $\phi$ defines the foliation and is fixed to unitary gauge, whereas the other scalar field $\psi$ remains explicit.

We consider a broad class of actions of the form~\eqref{eq:S0}, in which metric velocities enter through $K_{ij}$ and the explicit scalar $\psi$ appears with Lie derivatives up to $\pounds_{\bm{u}}^{2}\psi$, together with spatial derivatives. Our main task is to identify the conditions under which a subclass of these theories propagates four physical degrees of freedom.
To this end, we introduce auxiliary variables and rewrite the theory in the first-order form \eqref{model}, which makes its primary and secondary constraints accessible. We then perform a nonlinear Hamiltonian constraint analysis to determine the number of degrees of freedom.

Since the Hamiltonian constraint analysis is technically involved and the
conditions emerge only after several successive stages, we summarize the
principal results below for ease of reference:
\begin{itemize}
	\item For the general action \eqref{eq:S0}, the operator $\mathcal P^{\alpha\beta}$ defined in
\eqref{eq:matrix} collects the Poisson brackets between the primary and
secondary constraints. If this operator is nondegenerate, the general action
\eqref{eq:S0} propagates five physical DOFs: two tensor modes and three scalar-sector modes. One scalar
	mode is associated with the SCG sector, one is the expected mode of
	$\psi$, and the third is the unwanted higher-derivative mode associated
	with $\psi$.
	
	\item To eliminate the unwanted mode, $\mathcal P^{\alpha\beta}$ must
	be degenerate. In the branch where the lapse Hessian and the effective
	tensor block are invertible and $\mathcal P^{\alpha\beta}$ has a single
	local zero-mode direction, the degeneracy condition is expressed by
	\eqref{eq:degeneracy condition}, with
	$\mathcal{D}(\vec{x},\vec{y})$ defined in
	\eqref{eq:degeneracy condition function}. This condition generates the
	tertiary constraint \eqref{ter_cons}, but is not by itself sufficient
	to fully eliminate the unwanted mode.
    
	\item The additional consistency condition
	\eqref{eq:consistency condition}, with
	$\mathcal{F}(\vec{x},\vec{y})$ given in
	\eqref{eq:conscalFxpl}, is required to completely eliminate the
	unwanted mode. When both the degeneracy and consistency conditions are
	satisfied, the number of physical DOFs is reduced to four. The
	preservation of the tertiary constraint either closes without
	generating a further constraint or yields the quaternary constraint
	\eqref{eq:quaternary constraint}.
\end{itemize}

We also examine how our construction behaves under the two-field disformal transformation by deriving its invertibility criterion and expressing it in
unitary gauge. A generic transformation introduces a velocity of the lapse and hence maps an action of the form \eqref{eq:S0} outside the class analyzed
here. We therefore identify a restricted invertible subclass that remains within this framework. An invertible and regular transformation in this
subclass preserves the number of physical degrees of freedom, although the transformed action need not realize the particular degeneracy branch analyzed in this work.

The paper is organized as follows. In Sec. \ref{sec:scgsf}, we first formulate the spatially covariant scalar field theory and its first-order auxiliary representation. In Sec. \ref{sec:ham}, we carry out the Hamiltonian constraint analysis and show that the theory generally propagates five DOFs. In Sec. \ref{sec:dccond}, we derive the degeneracy and consistency conditions and show, in the branch specified above, that their simultaneous imposition reduces the number of degrees of freedom to four. We subsequently analyze multi-field disformal transformations in Sec. \ref{sec:dftrans}, including their invertibility and their action on the unitary-gauge variables. In Sec. \ref{sec:con} we summarize our results.

\section{Spatially covariant scalar field theory} \label{sec:scgsf}

Reference~\cite{Yu:2024sed} extended single-field SCG by assigning independent
foliations to two scalar fields, $\phi$ and $\psi$, and constructing
interactions from the geometric quantities associated with each foliation.
A brief review is given in Appendix~\ref{app:dec}.

In this work, we develop a complementary route toward ghost-free
bi-scalar-tensor theories with four physical DOFs. We let one scalar, $\phi$,
define a single foliation and work in its unitary gauge, while keeping $\psi$
as an explicit field. The resulting gauge-fixed Lagrangian takes the form of
SCG coupled to an additional scalar field, which we refer to as a
``spatially covariant scalar field theory.'' This formulation separates Lie
and spatial derivatives and is therefore well suited to controlling the
highest time derivatives and performing a nonlinear Hamiltonian analysis.
Temporal diffeomorphism invariance can subsequently be restored through the
Stueckelberg procedure, yielding a generally covariant bi-scalar-tensor
formulation.

Let $\Sigma_\phi$ denote a constant-$\phi$ hypersurface. We assume that the
gradient of $\phi$ is timelike and define
$X\coloneqq-\frac{1}{2}\nabla_a\phi\nabla^a\phi>0$. The future-directed
unit normal is $u_a\coloneqq-\nabla_a\phi/\sqrt{2X}$, and the induced
metric is $h_{ab}\coloneqq g_{ab}+u_a u_b$. We denote by $\mathrm{D}_a$ the
spatial covariant derivative compatible with $h_{ab}$ and define the
extrinsic curvature and acceleration by
$K_{ab}\coloneqq\frac{1}{2}\pounds_{\bm{u}}h_{ab}$ and
$a_a\coloneqq u^b\nabla_b u_a$, respectively. The first and second covariant
derivatives of $\psi$ decompose relative to $\Sigma_\phi$ as
\begin{align}
	\nabla_{a}\psi & =-u_{a}\pounds_{\bm{u}}\psi+\mathrm{D}_{a}\psi, \label{dec_nbl1}\\
	\nabla_{a}\nabla_{b}\psi & =\mathrm{D}_{a}\mathrm{D}_{b}\psi-K_{ab}\pounds_{\bm{u}}\psi-2u_{(a}\mathrm{D}_{b)}\pounds_{\bm{u}}\psi+2K^{c}{}_{(a}u_{b)}\mathrm{D}_{c}\psi+\left(\pounds^{2}_{\bm{u}}\psi-a^{c}\mathrm{D}_{c}\psi\right)u_{a}u_{b}. \label{dec_nbl2}
\end{align}
Throughout this paper, $a,b,c,\ldots$ are spacetime indices, whereas
$i,j,k,\ldots$ are spatial indices intrinsic to $\Sigma_\phi$.

The preceding $3+1$ decomposition is geometric and does not require a
coordinate choice. We now choose coordinates adapted to the foliation and use
the temporal coordinate freedom to set $\phi=t$, which we dub the unitary
gauge of $\phi$. It fixes only the time coordinate, while the spatial
coordinates $x^i$ on the hypersurfaces remain unfixed. Time-dependent spatial
diffeomorphisms therefore remain as residual gauge transformations. In this gauge,
$N=1/\sqrt{2X}$. In terms of the usual lapse and shift, the normal vector and
the spatial tensors take the component forms
\begin{equation}
	u_a=(-N,0_i),\qquad u^a=\frac{1}{N}(1,-N^i),
\qquad
	h_{ab}=\left(\begin{array}{cc}
		N_{i}N^{i} & N_{j}\\
		N_{i} & h_{ij}
	\end{array}\right),
\end{equation}
and
\begin{equation}
	K_{ab}=\left(\begin{array}{cc}
		K_{ij}N^{i}N^{j} & K_{ij}N^{i}\\
		K_{ij}N^{j} & K_{ij}
	\end{array}\right).
\end{equation}

In these adapted coordinates, the derivatives of $\psi$ in
(\ref{dec_nbl1}) and (\ref{dec_nbl2}) take the component forms
\begin{equation}
	\nabla_a\psi=
	\left(\begin{array}{c}
		N\pounds_{\bm{u}}\psi+N^i\mathrm{D}_i\psi\\
		\mathrm{D}_i\psi
	\end{array}\right), \label{eq:dec_nbl1_psi}
\end{equation}
and
\begin{equation}
	\nabla_a\nabla_b\psi=
	\left(\begin{array}{cc}
		N^2\varPsi+2NN^i\varPsi_i+N^iN^j\varPsi_{ij}
		& N\varPsi_j+N^i\varPsi_{ij}\\
		N\varPsi_i+N^j\varPsi_{ij} & \varPsi_{ij}
	\end{array}\right), \label{eq:dec_nbl2_psi}
\end{equation}
where we define
\begin{align}
	\varPsi&\coloneqq\pounds_{\bm{u}}^2\psi-a^k\mathrm{D}_k\psi,\\
	\varPsi_i&\coloneqq\mathrm{D}_i\pounds_{\bm{u}}\psi-K_i{}^k\mathrm{D}_k\psi,\\
	\varPsi_{ij}&\coloneqq\mathrm{D}_i\mathrm{D}_j\psi-K_{ij}\pounds_{\bm{u}}\psi,
\end{align}
which serve as shorthand quantities.

We are now ready to construct the action. Relative to single-field SCG, the
additional building blocks supplied by $\psi$ are $\psi$,
$\pounds_{\bm{u}}\psi$,
$\pounds^2_{\bm{u}}\psi$, and their spatial derivatives.
The metric sector is constructed from the usual single-field
SCG variables \citep{Gao:2014soa,Gao:2014fra}. We restrict its explicit
velocities to the extrinsic curvature $K_{ij}$, excluding Lie derivatives of
the lapse and of $K_{ij}$, which keeps the metric sector first order in time.
We then couple the second scalar $\psi$, allowing its first and second Lie
derivatives along the preferred normal. Arbitrary spatial covariant
derivatives of the admissible building blocks are allowed. A broad action
containing up to the second Lie derivative of $\psi$ can therefore be written
as
\begin{equation}
	S_{0}\equiv\int\mathrm{d}t\mathrm{d}^{3}x\,N\sqrt{h}\mathcal{L}_{0}\left(t,N,h_{ij},K_{ij},{}^{3}\!R_{ij};\psi,\pounds_{\bm{u}}\psi,\pounds^{2}_{\bm{u}}\psi;\mathrm{D}_{i}\right).\label{eq:S0}
\end{equation}
Equation~(\ref{eq:S0}) defines the broad class that is the starting point of
our analysis. We emphasize that it is not by itself assumed to propagate four DOFs. Because the
action contains $\pounds_{\bm{u}}^2\psi$, a generic nondegenerate member can
propagate an additional scalar mode. Our aim is to derive the degeneracy and
consistency conditions under which a subclass of Eq.~(\ref{eq:S0}) propagates
four physical DOFs.

Although $N^i$ is absent from the list of independent arguments of
$\mathcal{L}_0$, the action is not independent of the shift. Under
time-dependent spatial diffeomorphisms, $N^i$ transforms inhomogeneously as a
gauge connection and therefore cannot enter as an independent spatial tensor
without breaking the residual symmetry. In adapted coordinates, its allowed
dependence is encoded in covariant normal-evolution combinations, for example,
$\pounds_{\bm{u}}\psi=N^{-1}(\partial_t\psi-\pounds_{\vec{N}}\psi)$ and
$K_{ij}=(2N)^{-1}(\partial_t h_{ij}-\pounds_{\vec{N}}h_{ij})$, and similarly
in $\pounds_{\bm{u}}^2\psi$. Thus, the shift remains a gauge variable rather
than an additional physical building block.

The relation to the interacting-hypersurface construction can now be stated
more precisely. One first decomposes the geometric quantities associated with
$\Sigma_\psi$ with respect to $\Sigma_\phi$, as described in
Appendix~\ref{app:dec}, and then chooses coordinates adapted to $\Sigma_\phi$.
Those members whose resulting single-foliation expressions contain no metric
velocities beyond $K_{ij}$ and no Lie derivatives of $\psi$ beyond
$\pounds_{\bm{u}}^2\psi$ form a subclass of Eq.~(\ref{eq:S0}). As a concrete
illustration, consider the model (\ref{eq:model}) studied in
Ref.~\cite{Yu:2024sed}. Its expression in the unitary gauge of $\phi$, obtained
by decomposing all quantities with respect to $\Sigma_\phi$, is
\begin{align}
	\mathcal{L}= & \left(c_{1}-c_{3}\alpha\right)K_{ij}K^{ij}+c_{2}K^{2}+c_{4}{}^{3}\!R\nonumber \\
	& +c_{3}K^{ij}\Biggl\{\frac{1}{2\sqrt{2}}Y^{-\frac{3}{2}}\mathrm{D}_{j}Y\mathrm{D}_{i}\psi-\frac{1}{\sqrt{2Y}}\mathrm{D}_{i}\mathrm{D}_{j}\psi-\alpha^{2}a_{i}\left(\frac{1}{\sqrt{2Y}}\mathrm{D}_{j}\psi\right)+\frac{1}{4\sqrt{2}}\left(Y^{-\frac{5}{2}}\mathrm{D}_{k}Y\right)\mathrm{D}_{j}\psi\mathrm{D}^{k}\psi\mathrm{D}_{i}\psi\nonumber \\
	& -\left(\frac{1}{2Y}\right)^{\frac{3}{2}}\mathrm{D}_{j}\psi\mathrm{D}^{k}\psi\mathrm{D}_{k}\mathrm{D}_{i}\psi-\alpha\frac{1}{2Y}\mathrm{D}_{j}\psi\left(\mathrm{D}_{i}\left(\pounds_{\bm{u}}\psi\right)+\frac{\mathrm{D}_{i}N}{N}\pounds_{\bm{u}}\psi\right)\nonumber \\
	& +\frac{1}{4}\alpha\mathrm{D}_{i}\psi\mathrm{D}_{j}\psi Y^{-2}\left[\pounds_{\bm{u}}\psi\pounds^{2}_{\bm{u}}\psi-\mathrm{D}^{k}\psi\left(\mathrm{D}_{k}\left(\pounds_{\bm{u}}\psi\right)+\frac{\mathrm{D}_{k}N}{N}\pounds_{\bm{u}}\psi\right)+K^{kl}\mathrm{D}_{k}\psi\mathrm{D}_{l}\psi\right]\Biggr\}, \label{Laginths}
\end{align}
where\footnote{Throughout this paper, a ``$\coloneqq$'' denotes a definition, whereas a ``$\equiv$'' denotes an identity.}
\begin{align}
	Y&\coloneqq-\frac{1}{2}\nabla_a\psi\nabla^a\psi=\frac{1}{2}\left[(\pounds_{\bm{u}}\psi)^2-\mathrm{D}_i\psi \mathrm{D}^i\psi\right],
\\
	\alpha&\coloneqq u_a v^a=-\frac{\pounds_{\bm{u}}\psi}{\sqrt{2Y}},
\end{align}
and the coefficients $c_i$ ($i=1,2,3,4$) are understood to be functions of
$t$, $N$, $\psi$, $\pounds_{\bm{u}}\psi$, and spatial scalar combinations
constructed from $\mathrm{D}_i\psi$. Equation~(\ref{Laginths}) is therefore
an explicit example of the general action (\ref{eq:S0}).

In the subsequent sections, we determine the nonlinear constraint structure and the
number of propagating DOFs for the general action (\ref{eq:S0}). Directly
performing the Legendre transformation is impractical because the dependence
on the velocities can be nonlinear and the velocity-momentum relations need
not be invertible. We instead introduce independent auxiliary fields $A$,
$F$, and $B_{ij}$ for $\pounds_{\bm{u}}\psi$,
$\pounds_{\bm{u}}^2\psi$, and $K_{ij}$, respectively, together with
Lagrange multipliers $\varLambda$, $\tilde{\varLambda}$, and
$\varLambda^{ij}$. The action \eqref{eq:S0} can then be written in the first-order form
\begin{equation}
	\tilde{S}=S+\int\mathrm{d}t\mathrm{d}^{3}x\,N\sqrt{h}\left[\varLambda\left(\pounds_{\bm{u}}\psi-A\right)+\tilde{\varLambda}\left(\pounds_{\bm{u}}A-F\right)+\varLambda^{ij}\left(K_{ij}-B_{ij}\right)\right],\label{eq:Sbar}
\end{equation}
where
\begin{equation}
	S\equiv \int\mathrm{d}t\mathrm{d}^{3}x\,N\sqrt{h}\mathcal{L}\left(t,N,h_{ij},B_{ij},{}^{3}\!R_{ij};\psi,A,F;\mathrm{D}_{i}\right)\label{eq:S}
\end{equation}
denotes $S_{0}$ in (\ref{eq:S0}) after the replacements
$\pounds_{\bm{u}}\psi\rightarrow A$,
$\pounds^{2}_{\bm{u}}\psi\rightarrow F$, and
$K_{ij}\rightarrow B_{ij}$. Variations with respect to
$\varLambda$, $\tilde{\varLambda}$, and $\varLambda^{ij}$ impose these
three defining relations, whose substitution recovers $S_0$. Treating $A$,
$F$, and $B_{ij}$ as independent variables also prevents time derivatives of
the lapse and shift from appearing explicitly in this first-order action.

The multipliers in Eq.~(\ref{eq:Sbar}) can be related to variational
derivatives of $S$. Variations with respect to $F$ and $B_{ij}$ give the first
two relations below, while variation with respect to $A$, after integrations
by parts in time and space, gives the third:
\begin{align}
	\tilde{\varLambda} & =\frac{1}{N\sqrt{h}}\frac{\delta S}{\delta F},\qquad\varLambda^{ij}=\frac{1}{N\sqrt{h}}\frac{\delta S}{\delta B_{ij}},\label{eq:Lmdsol1}\\
	\varLambda & =\frac{1}{N\sqrt{h}}\frac{\delta S}{\delta A}-\frac{1}{N\sqrt{h}}\partial_{t}\left(\sqrt{h}\tilde{\varLambda}\right)+\frac{1}{N\sqrt{h}}\partial_{i}\left(\sqrt{h}\tilde{\varLambda}N^{i}\right).\label{eq:Lmdsol2}
\end{align}
Rather than substituting Eqs.~(\ref{eq:Lmdsol1}) and
(\ref{eq:Lmdsol2}) back into the action, we identify the coefficients of
$\dot\psi$, $\dot A$, and $\dot h_{ij}$ in Eq.~(\ref{eq:Sbar}) as the
corresponding conjugate momenta:
\begin{equation}
	p_{\psi}=\frac{\delta\tilde{S}}{\delta\dot{\psi}}=\sqrt{h}\varLambda,\quad
	p_{A}=\frac{\delta\tilde{S}}{\delta\dot{A}}=\sqrt{h}\tilde{\varLambda},\quad
	\pi^{ij}=\frac{\delta\tilde{S}}{\delta\dot{h}_{ij}}=\frac{1}{2}\sqrt{h}\varLambda^{ij}. \label{mom_Lagmtl}
\end{equation}
These relations define a change of variables from the multipliers to the
corresponding canonical momenta. In terms of the latter,
Eq.~(\ref{eq:Sbar}) becomes
\begin{equation}
	\tilde{S}\equiv S+\int\mathrm{d}t\mathrm{d}^{3}x\,N\left[p_{\psi}\left(\pounds_{\bm{u}}\psi-A\right)+p_{A}\left(\pounds_{\bm{u}}A-F\right)+2\pi^{ij}\left(K_{ij}-B_{ij}\right)\right],\label{model}
\end{equation}
which will be the starting point of the following analysis.
Equation~(\ref{model}) is an equivalent first-order Hamilton-Pontryagin
action in the $(\psi,p_\psi)$, $(A,p_A)$, and $(h_{ij},\pi^{ij})$ sectors.
In this form, the momenta are independent variables. The relations in
Eqs.~(\ref{eq:Lmdsol1}) and (\ref{eq:Lmdsol2}) reappear as equations of motion
obtained by varying $A$, $F$, and $B_{ij}$, whereas variations of the momenta
impose
$A=\pounds_{\bm{u}}\psi$, $F=\pounds_{\bm{u}}A$, and $B_{ij}=K_{ij}$,
whose substitution recovers $S_0$. Since no time derivatives of $N$, $N^i$,
$F$, or $B_{ij}$ occur in Eq.~(\ref{model}), their conjugate momenta vanish,
as shown in the next section.

\section{Hamiltonian constraint analysis} \label{sec:ham}

\subsection{Primary constraints and the Hamiltonian}

In this section, we perform a Hamiltonian constraint analysis for
the model (\ref{model}). Counting independent tensor components,
$\tilde{S}$ depends on 19 configuration variables,
\begin{equation}
	\varPhi_{I}=\left\{ N,N^{i},h_{ij},\psi,A,F,B_{ij}\right\} ,
\end{equation}
where $I,J,\ldots$ collectively label the different kinds of variables
and their independent components. Together with their conjugate momenta
$\varPi^{I}=\left\{ \pi_{N},\pi_{i},\pi^{ij},p_{\psi},p_{A},p_{F},p^{ij}\right\} $,
they span a 38-dimensional phase space at each spatial point. The conjugate momenta are defined
as 
\begin{equation}
	\varPi^{I}=\frac{\delta\tilde{S}}{\delta\dot{\varPhi}_{I}}. \label{momdef}
\end{equation}
For $h_{ij}$, $A$, and $\psi$, Eq.~\eqref{momdef} reproduces the momenta
introduced in Eq.~\eqref{model}. For the variables whose velocities are absent, we find
\begin{align}
	&\pi_{N}=\frac{\delta\tilde{S}}{\delta\dot{N}}=0,\qquad\pi_{i}=\frac{\delta\tilde{S}}{\delta\dot{N}^{i}}=0,
\\
	&p_{F}=\frac{\delta\tilde{S}}{\delta\dot{F}}=0,\qquad p^{ij}=\frac{\delta\tilde{S}}{\delta\dot{B_{ij}}}=0.
\end{align}
The explicit absence of time derivatives of $N,N^{i},F,B_{ij}$
yields 11 primary constraints $\varphi^{A}=\left\{ \pi_{N},\pi_{i},p_{F},p^{ij}\right\} $,
with\footnote{Here and throughout this paper, we use $A,B$ as indices for the primary constraints, which should not be confused with the auxiliary variable $A$.}
\begin{equation}
	\pi_{N}\approx0,\quad\pi_{i}\approx0,\quad p_{F}\approx0,\quad p^{ij}\approx0.
\end{equation}

The canonical Hamiltonian density obtained directly by the Legendre transformation is
\begin{equation}
	\mathcal{H}_{0}=  \dot{\psi}p_{\psi}+\dot{A}p_{A}+\dot{h}_{ij}\pi^{ij}-N\sqrt{h}\tilde{\mathcal{L}},
\end{equation}
where $\tilde{\mathcal{L}}$ denotes the Lagrangian
corresponding to the action $\tilde{S}$ in \eqref{model}.
After some manipulations, we find
\begin{align}
	\mathcal{H}_{0}=  & \left(N\pounds_{\bm{u}}\psi+\pounds_{\vec{N}}\psi\right)p_{\psi}+\left(N\pounds_{\bm{u}}A+\pounds_{\vec{N}}A\right)p_{A}+\left(2NK_{ij}+\pounds_{\vec{N}}h_{ij}\right)\pi^{ij}-N\sqrt{h}\tilde{\mathcal{L}}\nonumber \\
	= & N\left(Ap_{\psi}+Fp_{A}+2B_{ij}\pi^{ij}-\sqrt{h}\mathcal{L}\right)+p_{\psi}\pounds_{\vec{N}}\psi+p_{A}\pounds_{\vec{N}}A+\pi^{ij}\pounds_{\vec{N}}h_{ij}. \label{calH0}
\end{align}

Because the Legendre map is singular, the canonical Hamiltonian is fixed
only on the primary-constraint surface. Its extension away from that surface
is defined only up to terms proportional to primary constraints. Following
Ref.~\cite{Gao:2018znj}, we choose the extension that simplifies the
subsequent Poisson-bracket calculation,
\begin{equation}
	H_{0}\coloneqq\int\mathrm{d}^{3}x\left(NC\right)+X[\vec{N}], \label{Ham_canon}
\end{equation}
where
\begin{equation}
	C=Ap_{\psi}+Fp_{A}+2B_{ij}\pi^{ij}-\sqrt{h}\mathcal{L}.\label{eq:C}
\end{equation}
For a general spatial vector field $\vec{\xi}$, $X[\vec{\xi}]$ is
defined as 
\begin{align}
	X[\vec{\xi}] & \coloneqq\int\mathrm{d}^{3}x\left.\underset{I}{\sum}\varPi^{I}\pounds_{\vec{\xi}}\varPhi_{I}\right.\nonumber \\
	& =\int\mathrm{d}^{3}x\left(p_{\psi}\pounds_{\vec{\xi}}\psi+p_{A}\pounds_{\vec{\xi}}A+\pi^{ij}\pounds_{\vec{\xi}}h_{ij}+\pi_{N}\pounds_{\vec{\xi}}N+\pi_{i}\pounds_{\vec{\xi}}N^{i}+p_{F}\pounds_{\vec{\xi}}F+p^{ij}\pounds_{\vec{\xi}}B_{ij}\right). \label{Xdef}
\end{align}
The quantity $C$ in Eq.~\eqref{eq:C} contains $\mathcal{L}$, rather than
$\tilde{\mathcal{L}}$: the first-order terms that distinguish
$\tilde{\mathcal{L}}$ from $\mathcal{L}$ cancel the corresponding
velocity-momentum terms in the Legendre transformation. On the primary-constraint
surface, the Hamiltonian in Eq.~\eqref{Ham_canon} therefore agrees with
$\int \mathrm{d}^{3}x\,\mathcal{H}_{0}$ obtained from Eq.~\eqref{calH0}.
Their off-surface difference is proportional to the primary constraints.
Integrating Eq.~\eqref{Xdef} by parts and discarding a boundary term gives the
familiar form
\begin{align}
	H_{0} & \simeq \int\mathrm{d}^{3}x\left(NC+N^{i}\mathcal{C}_{i}\right),
\end{align}
where $C$ is given in Eq.~\eqref{eq:C}, and $\mathcal C_i$ is
\begin{align}
	\mathcal{C}_{i} ={}&p_{\psi}\mathrm{D}_{i}\psi+p_{A}\mathrm{D}_{i}A
	-2\sqrt{h}\mathrm{D}_{j}\left(\frac{\pi^{j}_{i}}{\sqrt{h}}\right)
	+\pi_{N}\mathrm{D}_{i}N+p_{F}\mathrm{D}_{i}F
	+p^{jk}\mathrm{D}_{i}B_{jk}
	-2\sqrt{h}\mathrm{D}_{j}\left(\frac{B_{ik}p^{jk}}{\sqrt{h}}\right)\nonumber\\
	&+\pi_{j}\mathrm{D}_{i}N^{j}
	+\sqrt{h}\mathrm{D}_{j}\left(\frac{\pi_{i}N^{j}}{\sqrt{h}}\right).\label{eq:calCi}
\end{align}
The last two terms in Eq.~\eqref{eq:calCi} follow from
$\pi_i\pounds_{\vec\xi}N^i$ in Eq.~\eqref{Xdef} and ensure that
$\mathcal C_i$ generates spatial diffeomorphisms also in the shift sector.
In contrast to GR, $C$ can generally depend on the lapse $N$.

The velocities in the directions associated with the primary constraints are
not fixed by the Legendre transformation. The most general Hamiltonian evolution
compatible with it is therefore generated by the total Hamiltonian,
\begin{equation}
	H_{\mathrm{T}} \coloneqq H_{0}+\int\mathrm{d}^{3}x\left(\lambda\pi_{N}+\lambda^{i}\pi_{i}+vp_{F}+v_{ij}p^{ij}\right). \label{Ham_total}
\end{equation}
Here $\lambda$, $\lambda^{i}$, $v$, and the symmetric tensor $v_{ij}$ are
a priori arbitrary functions of space and time that multiply, respectively,
$\pi_N$, $\pi_i$, $p_F$, and $p^{ij}$. Their preservation equations may either
determine some of these multipliers or generate further constraints.

\subsection{Secondary constraints}

For arbitrary functionals $\mathcal A$ and $\mathcal B$ of the canonical
variables $\{\varPhi_{I},\varPi^{I}\}$, the Poisson bracket is defined by
\begin{equation}
	\left[\mathcal A,\mathcal B\right]\equiv\sum_{I}\int\mathrm{d}^{3}x\left(\frac{\delta \mathcal A}{\delta\varPhi_{I}(\vec{x})}\frac{\delta \mathcal B}{\delta\varPi^{I}(\vec{x})}-\frac{\delta \mathcal A}{\delta\varPi^{I}(\vec{x})}\frac{\delta \mathcal B}{\delta\varPhi_{I}(\vec{x})}\right),
\end{equation}
and the time evolution of $\mathcal A$ is
\begin{equation}
	\frac{\mathrm{d}\mathcal A}{\mathrm{d}t}\approx\frac{\partial \mathcal A}{\partial t}+\left[\mathcal A,H_{\mathrm{T}}\right]. \label{timeevo_HamT}
\end{equation}

Constraints must be preserved in time.  Since the primary constraints
$\varphi^A$ have no explicit time dependence, their preservation equations
read
\begin{equation}
	\int\mathrm{d}^{3}y\left.\underset{B}{\sum}\left[\varphi^{A}\left(\vec{x}\right),\varphi^{B}\left(\vec{y}\right)\right]\lambda_{B}\left(\vec{y}\right)\right.+\left[\varphi^{A}\left(\vec{x}\right),H_{0}\right]\approx0,
\end{equation}
where $\lambda_{A}=\left\{ \lambda,\lambda^{i},v,v_{ij}\right\} $.
All primary-primary Poisson brackets vanish strongly, so this step does not
yet determine any multiplier. Secondary constraints arise if
$\left[\varphi^{A}(\vec{x}),H_{0}\right]\approx0$ yields relations among
the phase-space variables that are independent of the primary constraints. We have
\begin{equation}
	\left[\varphi^{A}\left(\vec{x}\right),H_{0}\right]=\int\mathrm{d}^{3}y\left[\varphi^{A}\left(\vec{x}\right),N\left(\vec{y}\right)C\left(\vec{y}\right)+N^{i}\left(\vec{y}\right)\mathcal{C}_{i}\left(\vec{y}\right)\right].
\end{equation}
For $\varphi^{A}=\left\{ \pi_{N},\pi_{i},p_{F},p^{ij}\right\} $, these
conditions, modulo primary constraints, read
\begin{equation}
	\left[\varphi^{A}\left(\vec{x}\right),H_{0}\right]\approx\left(\begin{array}{c}
		\mathcal{C}\left(\vec{x}\right)\\
		-\mathcal{C}_{i}\left(\vec{x}\right)\\
		\chi\left(\vec{x}\right)\\
		\chi^{ij}\left(\vec{x}\right)
	\end{array}\right)\approx 0,
\end{equation}
yielding 11 secondary constraints $\chi^{A}\coloneqq\left\{ \mathcal{C},\mathcal{C}_{i},\chi,\chi^{ij}\right\} $,
where 
\begin{align}
	\mathcal{C}\left(\vec{x}\right) & \coloneqq -C\left(\vec{x}\right)-\int\mathrm{d}^{3}y\left.N\left(\vec{y}\right)\frac{\delta C\left(\vec{y}\right)}{\delta N\left(\vec{x}\right)}\right.\nonumber \\
	& =-A\left(\vec{x}\right)p_{\psi}\left(\vec{x}\right)-F\left(\vec{x}\right)p_{A}\left(\vec{x}\right)-2B_{ij}\left(\vec{x}\right)\pi^{ij}\left(\vec{x}\right)+\frac{\delta S}{\delta N\left(\vec{x}\right)}\approx0,
\\
	\chi\left(\vec{x}\right) & \coloneqq -N\left(\vec{x}\right)p_{A}\left(\vec{x}\right)+\frac{\delta S}{\delta F\left(\vec{x}\right)}\approx0,\label{eq:=00003D00005Cchi(x)}
\\
	\chi^{ij}\left(\vec{x}\right)&\coloneqq-2N\left(\vec{x}\right)\pi^{ij}\left(\vec{x}\right)+\frac{\delta S}{\delta B_{ij}\left(\vec{x}\right)}\approx0,\label{eq:=00003D00005Cchi^ij(x)}
\end{align}
and $\mathcal{C}_{i}\approx0$ is defined in Eq.~\eqref{eq:calCi}.
If $C$ is independent of $N$, as is the case for the ADM Hamiltonian
constraint of GR, our sign convention gives $\mathcal C=-C$. Hence
$\mathcal C\approx0$ is equivalent to $C\approx0$. In the present theory,
$C$ generally depends on $N$, and the functional-derivative term in
$\mathcal C$ prevents this simple identification.

\subsection{Constraint algebra and number of DOFs}

\label{subsec:Constraint-algebra}

The secondary constraints must also be preserved in time, which gives
\begin{equation}
	\int\mathrm{d}^{3}y\left.\underset{B}{\sum}\left[\chi^{A}\left(\vec{x}\right),\varphi^{B}\left(\vec{y}\right)\right]\lambda_{B}\left(\vec{y}\right)\right.+\left[\chi^{A}\left(\vec{x}\right),H_{0}\right] + \frac{\partial \chi^{A}\left(\vec{x}\right)}{\partial t}\approx 0.\label{eq:tertiary constraints}
\end{equation}
Note that we include the term $\frac{\partial \chi^{A}\left(\vec{x}\right)}{\partial t}$ since the secondary constraints may depend explicitly on time through $\mathcal L(t,\ldots)$.
Whether these equations determine multipliers or generate tertiary constraints
is controlled in part by the Poisson brackets between the primary and secondary
constraint sets, summarized in Table~\ref{table:matrix}. Here and below, a zero
in a constraint table denotes a weakly vanishing bracket.
\begin{table}[H]
	\centering %
	\begin{tabular}{c|cccc}
		$\left[\bullet,\bullet \right]$  & $\pi_{i}\left(\vec{y}\right)$  & $\pi_{N}\left(\vec{y}\right)$  & $p_{F}\left(\vec{y}\right)$  & $p^{kl}\left(\vec{y}\right)$\tabularnewline
		\hline 
		$\mathcal{C}_{i}\left(\vec{x}\right)$  & 0  & 0  & 0  & 0\tabularnewline
		$\mathcal{C}\left(\vec{x}\right)$  & 0  & $\left[\mathcal{C}\left(\vec{x}\right),\pi_{N}\left(\vec{y}\right)\right]$  & $\left[\mathcal{C}\left(\vec{x}\right),p_{F}\left(\vec{y}\right)\right]$  & $\left[\mathcal{C}\left(\vec{x}\right),p^{kl}\left(\vec{y}\right)\right]$\tabularnewline
		$\chi\left(\vec{x}\right)$  & 0  & $\left[\chi\left(\vec{x}\right),\pi_{N}\left(\vec{y}\right)\right]$  & $\left[\chi\left(\vec{x}\right),p_{F}\left(\vec{y}\right)\right]$  & $\left[\chi\left(\vec{x}\right),p^{kl}\left(\vec{y}\right)\right]$\tabularnewline
		$\chi^{ij}\left(\vec{x}\right)$  & 0  & $\left[\chi^{ij}\left(\vec{x}\right),\pi_{N}\left(\vec{y}\right)\right]$  & $\left[\chi^{ij}\left(\vec{x}\right),p_{F}\left(\vec{y}\right)\right]$  & $\left[\chi^{ij}\left(\vec{x}\right),p^{kl}\left(\vec{y}\right)\right]$\tabularnewline
	\end{tabular}
	\caption{Poisson brackets between the primary and secondary constraint sets.}
	\label{table:matrix} 
\end{table}

The constraints $\mathcal C$, $\chi$, and $\chi^{ij}$ are independent of
$N^i$, so their brackets with $\pi_i$ vanish strongly. Although
$\mathcal C_i$ contains the shift-sector terms displayed in
Eq.~\eqref{eq:calCi}, its bracket with $\pi_i$ is proportional to $\pi_j$
and its spatial derivatives and therefore vanishes weakly. More generally,
$\mathcal C_i$ is the generator of spatial diffeomorphisms and satisfies
\cite{Gao:2018znj}
\begin{equation}
	\left[\mathcal{C}_{i}\left(\vec{x}\right),Q\left(\vec{y}\right)\right]\approx0
	\label{PB_calCi_Q}
\end{equation}
for any constraint $Q\approx0$ that transforms covariantly under spatial
diffeomorphisms, with tensor indices suppressed. To see this, introduce
phase-space-independent test functions $\xi^i$ and $f$, with the index
structure and density weights chosen so that
$Q[f]\coloneqq\int\mathrm d^3y\,fQ$ is invariant under spatial
diffeomorphisms. Using
$X[\vec\xi]\simeq\int\mathrm d^3x\,\xi^i\mathcal C_i$, one obtains
\begin{align}
	& \int\mathrm{d}^{3}x\mathrm{d}^{3}y\,
	\xi^{i}\left(\vec{x}\right)f\left(\vec{y}\right)
	\left[\mathcal{C}_{i}\left(\vec{x}\right),Q\left(\vec{y}\right)\right]
	\simeq \left[X\big[\vec{\xi}\big],Q[f]\right]
	= \int\mathrm{d}^{3}y\,Q\left(\vec{y}\right)
	\pounds_{\vec{\xi}}f\left(\vec{y}\right).
	\label{PB_calCi_Q_proven}
\end{align}
If $Q\approx0$, the right-hand side vanishes weakly. The arbitrariness of
the test functions then yields Eq.~\eqref{PB_calCi_Q}.
In particular, all brackets involving $\mathcal C_i$ or $\pi_i$ and any
constraint vanish weakly, which is the canonical manifestation of the
residual spatial-diffeomorphism invariance.

For other Poisson brackets, after some manipulations, we find
\begin{align}
	\left[\mathcal{C}\left(\vec{x}\right),\pi_{N}\left(\vec{y}\right)\right] & =\frac{\delta^{2}S}{\delta N\left(\vec{x}\right)\delta N\left(\vec{y}\right)},\label{PB:calCpiN}\\
	\left[\mathcal{C}\left(\vec{x}\right),p_{F}\left(\vec{y}\right)\right] & =-p_{A}\left(\vec{x}\right)\delta^{(3)}\left(\vec{x}-\vec{y}\right)+\frac{\delta^{2}S}{\delta N\left(\vec{x}\right)\delta F\left(\vec{y}\right)}, \label{PB:calCpF}\\
	\left[\mathcal{C}\left(\vec{x}\right),p^{kl}\left(\vec{y}\right)\right] & =-2\pi^{kl}\left(\vec{x}\right)\delta^{(3)}\left(\vec{x}-\vec{y}\right)+\frac{\delta^{2}S}{\delta N\left(\vec{x}\right)\delta B_{kl}\left(\vec{y}\right)},
\\
	\left[\chi\left(\vec{x}\right),\pi_{N}\left(\vec{y}\right)\right] & =-p_{A}\left(\vec{x}\right)\delta^{(3)}\left(\vec{x}-\vec{y}\right)+\frac{\delta^{2}S}{\delta F\left(\vec{x}\right)\delta N\left(\vec{y}\right)},\\
	\left[\chi\left(\vec{x}\right),p_{F}\left(\vec{y}\right)\right] & =\frac{\delta^{2}S}{\delta F\left(\vec{x}\right)\delta F\left(\vec{y}\right)},\\
	\left[\chi\left(\vec{x}\right),p^{kl}\left(\vec{y}\right)\right] & =\frac{\delta^{2}S}{\delta F\left(\vec{x}\right)\delta B_{kl}\left(\vec{y}\right)},
\\
	\left[\chi^{ij}\left(\vec{x}\right),\pi_{N}\left(\vec{y}\right)\right] & =-2\pi^{ij}\left(\vec{x}\right)\delta^{(3)}\left(\vec{x}-\vec{y}\right)+\frac{\delta^{2}S}{\delta B_{ij}\left(\vec{x}\right)\delta N\left(\vec{y}\right)},\\
	\left[\chi^{ij}\left(\vec{x}\right),p_{F}\left(\vec{y}\right)\right] & =\frac{\delta^{2}S}{\delta B_{ij}\left(\vec{x}\right)\delta F\left(\vec{y}\right)},\\
	\left[\chi^{ij}\left(\vec{x}\right),p^{kl}\left(\vec{y}\right)\right] & =\frac{\delta^{2}S}{\delta B_{ij}\left(\vec{x}\right)\delta B_{kl}\left(\vec{y}\right)}.
\end{align}

Before proceeding, we note that some of the Poisson brackets above may vanish in special cases. For example, evaluating (\ref{PB:calCpF}) explicitly yields
\begin{align}
	& \left[\mathcal{C}\left(\vec{x}\right),p_{F}\left(\vec{y}\right)\right]
	\nonumber \\
	= {}&
	\frac{1}{N\left(\vec{x}\right)}
	\chi\left(\vec{x}\right)
	\delta^{(3)}\left(\vec{x}-\vec{y}\right)
	+\frac{1}{N\left(\vec{x}\right)}
	\underset{n=1}{\sum}
	\sqrt{h\left(\vec{x}\right)}
	\left(-1\right)^{n}
	\partial_{x^{k_{1}}}
	\partial_{x^{k_{2}}}
	\cdots
	\partial_{x^{k_{n}}}
	N\left(\vec{x}\right)
	\frac{\partial\mathcal{L}\left(\vec{x}\right)}
	{\partial\left(
		\partial_{x^{k_{1}}}
		\partial_{x^{k_{2}}}
		\cdots
		\partial_{x^{k_{n}}}
		F\left(\vec{x}\right)
		\right)}
	\delta^{(3)}\left(\vec{x}-\vec{y}\right)
	\nonumber \\
	& -
	\sum_{n=1}
	\sqrt{h\left(\vec{y}\right)}
	\left(-1\right)^{n}
	\partial_{y^{k_{1}}}
	\partial_{y^{k_{2}}}
	\cdots
	\partial_{y^{k_{n}}}
	\delta^{(3)}\left(\vec{y}-\vec{x}\right)
	\frac{\partial\mathcal{L}\left(\vec{y}\right)}
	{\partial\left(
		\partial_{y^{k_{1}}}
		\partial_{y^{k_{2}}}
		\cdots
		\partial_{y^{k_{n}}}
		F\left(\vec{y}\right)
		\right)}
	\nonumber \\
	& -
	\sum_{n,m=0}
	\left(-1\right)^{n}
	\partial_{y^{k_{1}}}
	\cdots
	\partial_{y^{k_{n}}}
	\left(
	N\left(\vec{y}\right)
	\sqrt{h\left(\vec{y}\right)}
	\frac{\partial^{2}\mathcal{L}\left(\vec{y}\right)}
	{\partial\left(
		\partial_{y^{k_{1}}}
		\cdots
		\partial_{y^{k_{n}}}
		F\left(\vec{y}\right)
		\right)
		\partial\left(
		\partial_{y^{l_{1}}}
		\cdots
		\partial_{y^{l_{m}}}
		N\left(\vec{y}\right)
		\right)}
	\partial_{y^{l_{1}}}
	\cdots
	\partial_{y^{l_{m}}}
	\delta^{(3)}\left(\vec{y}-\vec{x}\right)
	\right). \label{PBcalCpFxpl}
\end{align}
Thus, if the Lagrangian does not depend on spatial derivatives of $F$
and there are no mixed terms between $F$ and $N$ or its spatial derivatives,
then
$\left[\mathcal{C}\left(\vec{x}\right),p_{F}\left(\vec{y}\right)\right]
\approx 0$
holds automatically.
Similar simplifications can occur for
$\left[\mathcal{C}\left(\vec{x}\right),p^{kl}\left(\vec{y}\right)\right]$,
$\left[\chi\left(\vec{x}\right),\pi_{N}\left(\vec{y}\right)\right]$,
and
$\left[\chi^{ij}\left(\vec{x}\right),\pi_{N}\left(\vec{y}\right)\right]$.
In such cases, it is much easier to determine whether the matrix is degenerate and to identify possible tertiary constraints.

Returning to Eq.~\eqref{eq:tertiary constraints}, the preservation equation
for $\mathcal C_i$ is automatically satisfied on the constraint surface and
generates no tertiary constraint. We refer to Appendix~\ref{app:cc_calCi} for the detailed calculation.

For other secondary constraints,
we define 
\begin{equation}
	\mathcal{P}^{\alpha\beta}\left(\vec{x},\vec{y}\right)\coloneqq\left(\begin{array}{ccc}
		\left[\mathcal{C}\left(\vec{x}\right),\pi_{N}\left(\vec{y}\right)\right] & \left[\mathcal{C}\left(\vec{x}\right),p_{F}\left(\vec{y}\right)\right] & \left[\mathcal{C}\left(\vec{x}\right),p^{kl}\left(\vec{y}\right)\right]\\
		\left[\chi\left(\vec{x}\right),\pi_{N}\left(\vec{y}\right)\right] & \left[\chi\left(\vec{x}\right),p_{F}\left(\vec{y}\right)\right] & \left[\chi\left(\vec{x}\right),p^{kl}\left(\vec{y}\right)\right]\\
		\left[\chi^{ij}\left(\vec{x}\right),\pi_{N}\left(\vec{y}\right)\right] & \left[\chi^{ij}\left(\vec{x}\right),p_{F}\left(\vec{y}\right)\right] & \left[\chi^{ij}\left(\vec{x}\right),p^{kl}\left(\vec{y}\right)\right]
	\end{array}\right),\qquad\chi^{\alpha}\left(\vec{x}\right)\coloneqq\left(\begin{array}{c}
		\mathcal{C}\left(\vec{x}\right)\\
		\chi\left(\vec{x}\right)\\
		\chi^{ij}\left(\vec{x}\right)
	\end{array}\right),\label{eq:matrix}
\end{equation}
so Eq.~\eqref{eq:tertiary constraints} reduces to
\begin{equation}
	\int\mathrm{d}^{3}y\left.\mathcal{P}^{\alpha\beta}\left(\vec{x},\vec{y}\right)\lambda_{\beta}\left(\vec{y}\right)\right.+\left[\chi^{\alpha}\left(\vec{x}\right),H_{0}\right] + \frac{\partial \chi^{\alpha}\left(\vec{x}\right)}{\partial t} \approx 0,\label{eq:tertiary conditions of rest}
\end{equation}
where $\lambda_{\beta}=\left\{ \lambda,v,v_{ij}\right\} $ are the
corresponding multipliers. In the generic nondegenerate branch,
$\mathcal P^{\alpha\beta}$ is invertible as an integral-kernel operator
on the chosen function space, subject to the adopted boundary conditions.
Equation~\eqref{eq:tertiary conditions of rest}
then determines all $\lambda_\beta$ and produces no tertiary constraint.

Without additional degeneracy conditions, the algorithm therefore yields
the 22 constraints
$\{\pi_{N},\pi_{i},p_{F},p^{ij},\mathcal{C},\mathcal{C}_{i},\chi,\chi^{ij}\}
\approx0$. The six constraints $\pi_i$ and $\mathcal C_i$ are first-class.
The remaining 16 are second-class: the invertibility of the
primary-secondary block $\mathcal P^{\alpha\beta}$ makes the corresponding
16-component constraint-bracket operator nonsingular, independently of the
secondary-secondary block. Their brackets are summarized in
Table~\ref{table: Poisson brackets among primary and secondary constraints},
where ``$X$'' denotes a bracket that is generally nonvanishing. We use
``constraint matrix'' (or ``Dirac matrix'') for the integral-kernel operator
formed by these brackets. In the construction of Dirac brackets, its
invertible restriction to the second-class sector is understood.
\begin{center}
	\begin{table}[h]
		\begin{centering}
			\begin{tabular}{c|cccc|cccc}
				$\left[\bullet,\bullet\right]$  & $\pi_{i}\left(\vec{y}\right)$  & $\pi_{N}\left(\vec{y}\right)$  & $p_{F}\left(\vec{y}\right)$  & $p^{kl}\left(\vec{y}\right)$  & $\mathcal{C}_{i}\left(\vec{y}\right)$  & $\mathcal{C}\left(\vec{y}\right)$  & $\chi\left(\vec{y}\right)$  & $\chi^{ij}\left(\vec{y}\right)$\tabularnewline
				\hline 
				$\pi_{i}\left(\vec{x}\right)$  & 0  & 0  & 0  & 0  & 0  & 0  & 0  & 0\tabularnewline
				$\pi_{N}\left(\vec{x}\right)$  & 0  & 0  & 0  & 0  & 0  & $X$  & $X$  & $X$\tabularnewline
				$p_{F}\left(\vec{x}\right)$  & 0  & 0  & 0  & 0  & 0  & $X$  & $X$  & $X$\tabularnewline
				$p^{kl}\left(\vec{x}\right)$  & 0  & 0  & 0  & 0  & 0  & $X$  & $X$  & $X$\tabularnewline
				\hline 
				$\mathcal{C}_{i}\left(\vec{x}\right)$  & 0  & 0  & 0  & 0  & 0  & 0  & 0  & 0\tabularnewline
				$\mathcal{C}\left(\vec{x}\right)$  & 0  & $X$  & $X$  & $X$  & 0  & $X$  & $X$  & $X$\tabularnewline
				$\chi\left(\vec{x}\right)$  & 0  & $X$  & $X$  & $X$  & 0  & $X$  & $X$  & $X$\tabularnewline
				$\chi^{ij}\left(\vec{x}\right)$  & 0  & $X$  & $X$  & $X$  & 0  & $X$  & $X$  & $X$\tabularnewline
			\end{tabular}
			\par\end{centering}
		\caption{Poisson brackets among all primary and secondary constraints in the generic nondegenerate branch, before imposing any additional degeneracy condition.}
		\label{table: Poisson brackets among primary and secondary constraints} 
	\end{table}
	\par\end{center}

The Dirac-Bergmann counting formula therefore gives
\begin{align}
	\#_{\mathrm{DOF}} & =\frac{1}{2}\left(2\times\#_{\mathrm{var}}-2\times\#_{1\mathrm{st}}-\#_{2\mathrm{nd}}\right)\nonumber \\
	& =\frac{1}{2}\left(2\times19-2\times6-16\right)\nonumber \\
	& =5.
\end{align}
The five modes comprise the two tensor polarizations and three scalar modes.
One scalar mode resides in the spatially covariant metric sector and is
identified with the $\phi$ mode after temporal diffeomorphism invariance is
restored. The second is the expected mode carried by $\psi$. Because the
$\psi$ sector is nondegenerate and contains second Lie derivatives, it also
propagates a third scalar (the would-be Ostrogradsky mode associated with
$\psi$), which the conditions in the next section are designed to remove.

\section{Degeneracy and consistency conditions} \label{sec:dccond}

In this section, we seek the conditions required to eliminate the unwanted scalar mode.
Since the action in Eq.~\eqref{eq:S} may depend explicitly on time, it is
useful to introduce the multiplier-independent evolution operator
\begin{equation}
	\mathfrak{D}_{0}Q
	\coloneqq
	\frac{\partial Q}{\partial t}+[Q,H_{0}].
\end{equation}
According to Eq.~\eqref{timeevo_HamT}, the full time evolution is
\begin{equation}
	\frac{\mathrm{d}Q}{\mathrm{d}t}
	\approx
	\mathfrak{D}_{0}Q
	+\sum_A\int\mathrm{d}^{3}y\,
	\lambda_A(\vec y)[Q,\varphi^A(\vec y)].
\end{equation}
Thus, $\mathfrak{D}_{0}$ contains only the part independent of the
primary-constraint multipliers.

\subsection{The degeneracy condition}

To remove the unwanted mode, the operator
$\mathcal{P}^{\alpha\beta}(\vec{x},\vec{y})$ defined in
Eq.~\eqref{eq:matrix} must be degenerate so that the preservation equations for
the secondary constraints can yield an additional constraint rather than
determine all the corresponding multipliers. In this work, we focus on the branch in which
$\mathcal P^{\alpha\beta}$ has exactly one local zero-mode direction. In this case, there is a nontrivial left zero mode
$V_{\alpha}\coloneqq(U,V,V_{ij})$ satisfying
\begin{equation}
	\int\mathrm{d}^{3}x\,
	V_{\alpha}(\vec{x})
	\mathcal{P}^{\alpha\beta}(\vec{x},\vec{y})\approx 0,
	\label{eq:null vector}
\end{equation}
or, explicitly,
\begin{align}
	\int\mathrm{d}^{3}x\Bigl\{
	U(\vec{x})[\mathcal{C}(\vec{x}),\pi_{N}(\vec{y})]
	+V(\vec{x})[\chi(\vec{x}),\pi_{N}(\vec{y})]
	+V_{ij}(\vec{x})[\chi^{ij}(\vec{x}),\pi_{N}(\vec{y})]
	\Bigr\}& \approx 0,
	\label{eq:null vector of c}\\
	\int\mathrm{d}^{3}x\Bigl\{
	U(\vec{x})[\mathcal{C}(\vec{x}),p_{F}(\vec{y})]
	+V(\vec{x})[\chi(\vec{x}),p_{F}(\vec{y})]
	+V_{ij}(\vec{x})[\chi^{ij}(\vec{x}),p_{F}(\vec{y})]
	\Bigr\}& \approx 0,
	\label{eq:null vector of =00003D00005Cchi}\\
	\int\mathrm{d}^{3}x\Bigl\{
	U(\vec{x})[\mathcal{C}(\vec{x}),p^{kl}(\vec{y})]
	+V(\vec{x})[\chi(\vec{x}),p^{kl}(\vec{y})]
	+V_{ij}(\vec{x})[\chi^{ij}(\vec{x}),p^{kl}(\vec{y})]
	\Bigr\}& \approx 0.
	\label{eq:null vector of =00003D00005Cchi^ij}
\end{align}
Our task is to determine when these equations admit such a local zero-mode
family. We follow the method developed in Ref.~\cite{Gao:2018znj}.

In order to proceed, we select a subbranch in which 
$[\mathcal C(\vec x),\pi_N(\vec y)]$ is invertible. Let us introduce the Green kernel
$\mathcal I$ as its inverse,
\begin{equation}
	\int\mathrm{d}^{3}y\,
	\mathcal{I}(\vec{z},\vec{y})
	\frac{\delta^{2}S}{\delta N(\vec{x})\delta N(\vec{y})}
	\equiv-\delta^{(3)}(\vec{x}-\vec{z}),
	\label{eq:F(x.y)}
\end{equation}
or, using Eq.~\eqref{PB:calCpiN},
\begin{equation}
	\int\mathrm{d}^{3}y\,
	\mathcal{I}(\vec{z},\vec{y})
	[\mathcal{C}(\vec{x}),\pi_{N}(\vec{y})]
	\equiv-\delta^{(3)}(\vec{x}-\vec{z}).
\end{equation}
The minus sign in this definition is a convention that simplifies the formulas
below. Multiplying Eq.~\eqref{eq:null vector of c} by $\mathcal I$ then gives 
\begin{align}
	U(\vec{x}) = {}&\int\mathrm{d}^{3}z\mathrm{d}^{3}y\,
	\mathcal{I}(\vec{x},\vec{y})
	\Bigl\{V(\vec{z})[\chi(\vec{z}),\pi_{N}(\vec{y})]
	+V_{ij}(\vec{z})[\chi^{ij}(\vec{z}),\pi_{N}(\vec{y})]\Bigr\}.
	\label{eq:U in terms of V.V_ij}
\end{align}
Note that since \eqref{eq:null vector} holds only weakly,  the solution for $U$ contains ambiguity off the constraint surface. Here, we choose the strong equality for the solution for $U$, which can be understood as a ``representative'' solution of $U$ that holds in the whole phase space.
At this point, note that the existence of $\mathcal I$ by itself is not guaranteed for the general action
\eqref{eq:S0}. If $[\mathcal C(\vec x),\pi_N(\vec y)]$ is not invertible, one
must choose a different subbranch to proceed with the analysis.

Substituting Eq.~\eqref{eq:U in terms of V.V_ij} into
Eqs.~\eqref{eq:null vector of =00003D00005Cchi} and
\eqref{eq:null vector of =00003D00005Cchi^ij} yields
\begin{align}
	\int\mathrm{d}^{3}x\left[
	V(\vec{x})\mathcal{X}(\vec{x},\vec{y})
	+V_{ij}(\vec{x})\mathcal{G}^{ij}(\vec{x},\vec{y})\right]& \approx 0,
	\label{eq:degenerate condition of V and V_ij-1}\\
	\int\mathrm{d}^{3}x\left[
	V(\vec{x})\mathcal{Y}^{kl}(\vec{x},\vec{y})
	+V_{ij}(\vec{x})\mathcal{J}^{ij,kl}(\vec{x},\vec{y})\right]& \approx 0,
	\label{eq:degenerate condition of V and V_ij-2}
\end{align}
where 
\begin{align}
	\mathcal{X}(\vec{x},\vec{y})\coloneqq{}&
	[\chi(\vec{x}),p_{F}(\vec{y})]
	+\int\mathrm{d}^{3}z\mathrm{d}^{3}y'\,
	[\mathcal{C}(\vec{z}),p_{F}(\vec{y})]
	\mathcal{I}(\vec{z},\vec{y}')
	[\chi(\vec{x}),\pi_{N}(\vec{y}')],\\
	\mathcal{G}^{ij}(\vec{x},\vec{y})\coloneqq{}&
	[\chi^{ij}(\vec{x}),p_{F}(\vec{y})]
	+\int\mathrm{d}^{3}z\mathrm{d}^{3}y'\,
	[\mathcal{C}(\vec{z}),p_{F}(\vec{y})]
	\mathcal{I}(\vec{z},\vec{y}')
	[\chi^{ij}(\vec{x}),\pi_{N}(\vec{y}')],\\
	\mathcal{Y}^{kl}(\vec{x},\vec{y})\coloneqq{}&
	[\chi(\vec{x}),p^{kl}(\vec{y})]
	+\int\mathrm{d}^{3}z\mathrm{d}^{3}y'\,
	[\mathcal{C}(\vec{z}),p^{kl}(\vec{y})]
	\mathcal{I}(\vec{z},\vec{y}')
	[\chi(\vec{x}),\pi_{N}(\vec{y}')],\\
	\mathcal{J}^{ij,kl}(\vec{x},\vec{y})\coloneqq{}&
	[\chi^{ij}(\vec{x}),p^{kl}(\vec{y})]
	+\int\mathrm{d}^{3}z\mathrm{d}^{3}y'\,
	[\mathcal{C}(\vec{z}),p^{kl}(\vec{y})]
	\mathcal{I}(\vec{z},\vec{y}')
	[\chi^{ij}(\vec{x}),\pi_{N}(\vec{y}')]. \label{calJ}
\end{align}

We further restrict to the subbranch in which the effective tensor block
$\mathcal J^{ij,kl}$ is invertible on symmetric spatial tensors. This means
that, after the lapse direction has been eliminated, the six $B_{ij}$
directions contain no further zero mode and can be paired with their secondary
constraints. A singular $\mathcal J$ would signal an additional degeneracy in
this sector and requires a separate constraint analysis. On the present
subbranch, its inverse satisfies
\begin{equation}
	\int\mathrm{d}^{3}y\,
	(\mathcal{J}^{-1})_{mn,kl}(\vec{z},\vec{y})
	\mathcal{J}^{ij,kl}(\vec{x},\vec{y})
	\equiv \boldsymbol{1}^{ij}_{mn}\delta^{(3)}(\vec{x}-\vec{z}),
\end{equation}
where $\boldsymbol{1}^{ij}_{mn}\coloneqq
\delta^{(i}_{m}\delta^{j)}_{n}
=\frac12(\delta^i_m\delta^j_n+\delta^i_n\delta^j_m)$ is the identity on
symmetric rank-two tensors. Equation~\eqref{eq:degenerate condition of V and V_ij-2}
then determines 
\begin{equation}
	V_{ij}(\vec{x}) =
	\int\mathrm{d}^{3}z\,V(\vec{z})
	\mathcal{V}_{ij}(\vec{z},\vec{x}),
	\label{eq:V_ij}
\end{equation}
with
\begin{equation}
	\mathcal{V}_{ij}(\vec{z},\vec{x})\coloneqq
	-\int\mathrm{d}^{3}y\,
	(\mathcal{J}^{-1})_{ij,kl}(\vec{x},\vec{y})
	\mathcal{Y}^{kl}(\vec{z},\vec{y}).
	\label{eq:V_ij explicitlt}
\end{equation}
No assumption that $V$ is pointwise nonzero was used here. If
$V\equiv0$, invertibility of $\mathcal J$ forces $V_{ij}\equiv0$, and
Eq.~\eqref{eq:U in terms of V.V_ij} then gives $U\equiv0$. Hence the entire
zero mode is trivial. A nontrivial zero mode in this branch must therefore have
$V\not\equiv0$, while individual smearing functions may of course vanish at
some points.

Substitution of Eq.~\eqref{eq:V_ij} into
Eq.~\eqref{eq:degenerate condition of V and V_ij-1} gives
\begin{equation}
	\int\mathrm{d}^{3}x\,V(\vec{x})
	\mathcal{D}(\vec{x},\vec{y}) \approx 0,
	\label{eq:degeneracy equation}
\end{equation}
where 
\begin{equation}
	\boxed{\mathcal{D}(\vec{x},\vec{y})\coloneqq
	\mathcal{X}(\vec{x},\vec{y})
	-\int\mathrm{d}^{3}z\mathrm{d}^{3}y'\,
	\mathcal{G}^{ij}(\vec{z},\vec{y})
	(\mathcal{J}^{-1})_{ij,kl}(\vec{z},\vec{y}')
	\mathcal{Y}^{kl}(\vec{x},\vec{y}').}
	\label{eq:degeneracy condition function}
\end{equation}
Likewise, Eqs.~\eqref{eq:V_ij} and
\eqref{eq:U in terms of V.V_ij} give 
\begin{equation}
	U(\vec{x}) =
	\int\mathrm{d}^{3}z\,V(\vec{z})
	\mathcal{U}(\vec{z},\vec{x}),
	\label{eq:U}
\end{equation}
with
\begin{align}
	\mathcal{U}(\vec{z},\vec{x})\coloneqq{}&
	\int\mathrm{d}^{3}y\,
	\mathcal{I}(\vec{x},\vec{y})
	[\chi(\vec{z}),\pi_{N}(\vec{y})]
    +\int\mathrm{d}^{3}z'\mathrm{d}^{3}y\,
	\mathcal{V}_{ij}(\vec{z},\vec{z}')
	\mathcal{I}(\vec{x},\vec{y})
	[\chi^{ij}(\vec{z}'),\pi_{N}(\vec{y})].
	\label{eq:U explicitly}
\end{align}

A single isolated solution of Eq.~\eqref{eq:degeneracy equation} would provide
only a global functional zero mode. Eliminating one local canonical direction
requires the zero-mode family to be generated by an arbitrary smearing
$V(\vec x)$. Therefore, the reduced kernel must vanish on the existing
constraint surface,
\begin{equation}
	\boxed{\mathcal{D}(\vec{x},\vec{y})\approx0.}
	\label{eq:degeneracy condition}
\end{equation}
We use weak equality here because
$\mathcal{D}$ contains momenta through the Poisson brackets. After using
Eqs.~\eqref{eq:=00003D00005Cchi(x)} and
\eqref{eq:=00003D00005Cchi^ij(x)} to eliminate $p_A$ and $\pi^{ij}$, the
resulting expression is imposed as a functional identity on the Lagrangian in
Eq.~\eqref{eq:S0}.

\subsection{The additional tertiary constraint}

The tertiary constraint can be obtained directly by projecting the
preservation equations with the left zero mode. It is nevertheless useful to
replace the original constraints by combinations adapted to the left and right
zero-mode directions: this is an invertible change of constraint basis on the
branch under consideration and exposes the rank of the Dirac operator and its
first-class directions. No new constraint is introduced by the redefinition.

For the secondary constraints
$\chi^{\alpha}=\{\mathcal{C},\chi,\chi^{ij}\}$,
Eqs.~\eqref{eq:V_ij} and \eqref{eq:U} give 
\begin{align}
	\int\mathrm{d}^{3}x\,
	V_{\alpha}(\vec{x})\chi^{\alpha}(\vec{x})
	&=\int\mathrm{d}^{3}x\left[
	U(\vec{x})\mathcal{C}(\vec{x})
	+V(\vec{x})\chi(\vec{x})
	+V_{ij}(\vec{x})\chi^{ij}(\vec{x})\right]\nonumber\\
	&\equiv \int\mathrm{d}^{3}x\,
	V(\vec{x})\bar{\chi}(\vec{x})\approx0,
	\label{eq:redef_barchi}
\end{align}
where
\begin{equation}
	\bar{\chi}(\vec{x})\coloneqq
	\chi(\vec{x})+
	\int\mathrm{d}^{3}y\left[
	\mathcal{U}(\vec{x},\vec{y})\mathcal{C}(\vec{y})
	+\mathcal{V}_{ij}(\vec{x},\vec{y})\chi^{ij}(\vec{y})\right]
	\approx0.
	\label{eq:barchi_redef}
\end{equation}
This is a linear combination in the standard constraint-theory sense: the
coefficients may be phase-space-dependent functions or integral kernels. Their
Poisson brackets generate only terms proportional to the original constraints,
which vanish weakly, and the transformation is admissible provided its kernel
map is invertible on the chosen branch.

The operator $\mathcal P^{\alpha\beta}$ is formally self-adjoint,
\begin{equation}
	\mathcal{P}^{\alpha\beta}(\vec{x},\vec{y})
	=\mathcal{P}^{\beta\alpha}(\vec{y},\vec{x}),
\end{equation}
as follows from its explicit entries and integrations by parts under the same
boundary conditions. Hence it also has a right zero mode
$\tilde V_{\beta}\coloneqq(\tilde U,\tilde V,\tilde V_{ij})$ satisfying 
\begin{equation}
	\int\mathrm{d}^{3}y\,
	\mathcal{P}^{\alpha\beta}(\vec{x},\vec{y})
	\tilde{V}_{\beta}(\vec{y}) \approx 0.
	\label{eq:rightnvdef}
\end{equation}
Indeed,
\begin{equation}
	\int\mathrm{d}^{3}x\,
	V_{\alpha}(\vec{x})
	\mathcal{P}^{\alpha\beta}(\vec{x},\vec{y})
	=
	\int\mathrm{d}^{3}x\,
	\mathcal{P}^{\beta\alpha}(\vec{y},\vec{x})
	V_{\alpha}(\vec{x})\approx 0.
	\label{rightnv}
\end{equation}
Thus, the left and right kernels have the same generator. The smearing
functions $V$ and $\tilde V$ remain independent.

The right zero mode can be written as 
\begin{equation}
	\tilde{V}_{ij}(\vec{x})=
	\int\mathrm{d}^{3}z\,
	\tilde{V}(\vec{z})\tilde{\mathcal{V}}_{ij}(\vec{z},\vec{x}),
	\qquad
	\tilde{U}(\vec{x})=
	\int\mathrm{d}^{3}z\,
	\tilde{V}(\vec{z})\tilde{\mathcal{U}}(\vec{z},\vec{x}),
	\label{eq:V_ij~U~}
\end{equation}
where
\begin{equation}
	\tilde{\mathcal{V}}_{kl}(\vec{z},\vec{x})
	\coloneqq -\int\mathrm{d}^{3}y\,
	(\mathcal{J}^{-1})_{ij,kl}(\vec{y},\vec{x})
	\mathcal{G}^{ij}(\vec{y},\vec{z}),
\end{equation}
and
\begin{align}
	\tilde{\mathcal{U}}(\vec{z},\vec{x})\coloneqq{}&
	\int\mathrm{d}^{3}y\,
	\mathcal{I}(\vec{x},\vec{y})
	[\mathcal{C}(\vec{y}),p_{F}(\vec{z})]\nonumber\\
	&+\int\mathrm{d}^{3}z'\mathrm{d}^{3}y\,
	\tilde{\mathcal{V}}_{kl}(\vec{z},\vec{z}')
	\mathcal{I}(\vec{x},\vec{y})
	[\mathcal{C}(\vec{y}),p^{kl}(\vec{z}')].
\end{align}
For the primary constraints
$\varphi^{\alpha}=\{\pi_{N},p_{F},p^{ij}\}$, one then obtains 
\begin{align}
	\int\mathrm{d}^{3}x\,
	\tilde{V}_{\alpha}(\vec{x})\varphi^{\alpha}(\vec{x})
	&=\int\mathrm{d}^{3}x\left[
	\tilde{U}(\vec{x})\pi_{N}(\vec{x})
	+\tilde{V}(\vec{x})p_{F}(\vec{x})
	+\tilde{V}_{ij}(\vec{x})p^{ij}(\vec{x})\right]\nonumber\\
	&\equiv \int\mathrm{d}^{3}x\,
	\tilde{V}(\vec{x})\bar{p}_{F}(\vec{x})\approx0,
	\label{eq:redefinition equation of p_F}
\end{align}
with
\begin{equation}
	\bar{p}_{F}(\vec{x})\coloneqq
	p_{F}(\vec{x})+
	\int\mathrm{d}^{3}y\left[
	\tilde{\mathcal{U}}(\vec{x},\vec{y})\pi_{N}(\vec{y})
	+\tilde{\mathcal{V}}_{ij}(\vec{x},\vec{y})p^{ij}(\vec{y})\right]
	\approx0.
	\label{eq:redefine p_F}
\end{equation}
We may therefore use
\begin{equation}
	\left\{
	\pi_{i},\pi_{N},\bar{p}_{F},p^{ij},
	\mathcal{C}_{i},\mathcal{C},\bar{\chi},\chi^{ij}
	\right\}
	\label{eq:newsetcons}
\end{equation}
as an equivalent complete set of primary and secondary constraints.

The preservation equation for $\chi^\alpha$ is
$\mathfrak D_0\chi^\alpha+\int\mathcal P^{\alpha\beta}\lambda_\beta\approx0$.
Projecting it with the left zero mode removes all multipliers and gives
\begin{align}
	0&\approx\int\mathrm{d}^{3}x\,
	V_{\alpha}(\vec{x})\mathfrak D_{0}\chi^{\alpha}(\vec{x})
	\approx\mathfrak D_{0}
	\int\mathrm{d}^{3}x\,
	V_{\alpha}(\vec{x})\chi^{\alpha}(\vec{x})\nonumber\\
	&\equiv\mathfrak D_{0}
	\int\mathrm{d}^{3}x\,V(\vec{x})\bar\chi(\vec{x})
	\approx\int\mathrm{d}^{3}x\,V(\vec{x})\theta(\vec{x}),
	\label{eq:new constraint}
\end{align}
where the external smearing function $V$ is held fixed, while terms in
which $\mathfrak{D}_{0}$ acts on the phase-space-dependent zero-mode
coefficients are proportional to existing constraints and hence vanish
weakly. Since $V$ is an
arbitrary smearing function, a nontrivial projected equation defines the
tertiary constraint
\begin{equation}
	\boxed{\theta(\vec{x})\coloneqq
	\mathfrak D_{0}\bar{\chi}(\vec{x})\approx0.} 
    \label{ter_cons}
\end{equation}
Under the assumed corank-one and maximal-rank conditions, this is the only
independent tertiary constraint. All directions orthogonal to the left zero
mode determine the seven multipliers associated with $\pi_N$ and $p^{ij}$.
If the projected expression is identically dependent on the existing
constraints, the constraint chain and degree-of-freedom count change, which may
signal an additional gauge degeneracy but must be classified separately. Note that the preservation of $\theta$ must also be examined. We do so in the next subsection.

The zero-mode-adapted basis makes the vanishing brackets transparent. For
$\bar\chi$, Eqs.~\eqref{eq:redef_barchi} and \eqref{eq:null vector} imply
\begin{align}
	\left[
	\int\mathrm{d}^{3}x\,V(\vec{x})\bar\chi(\vec{x}),
	\varphi^{\beta}(\vec{y})
	\right]
	&\approx\int\mathrm{d}^{3}x\,
	V_{\alpha}(\vec{x})
	\mathcal P^{\alpha\beta}(\vec{x},\vec{y}) \approx 0,
\end{align}
and hence
\begin{equation}
	[\bar\chi(\vec{x}),\varphi^{\beta}(\vec{y})]\approx0.
\end{equation}
Similarly, the mutual brackets of the original primary constraints vanish,
and the right-zero-mode relation gives
\begin{equation}
	[\bar p_F(\vec{x}),\varphi^{\beta}(\vec{y})]\approx0,
	\qquad
	[\bar p_F(\vec{x}),\chi^{\beta}(\vec{y})]\approx0.
    \label{eq:pf-varphi-pf-chi}
\end{equation}
Consequently,
\begin{equation}
	[\bar p_F(\vec{x}),\bar p_F(\vec{y})]\approx0,
	\qquad
	[\bar p_F(\vec{x}),\bar\chi(\vec{y})]\approx0.
\end{equation}
These weak equalities include the terms generated by Poisson brackets of the
phase-space-dependent kernel coefficients, because such terms multiply
existing constraints.

Before imposing any further condition, no additional weakly vanishing bracket
between $\theta$ and the non-spatial constraint sector follows from the
zero-mode construction. The 11 primary constraints
$\{\pi_i,\pi_N,\bar p_F,p^{ij}\}$, 11 secondary constraints
$\{\mathcal C_i,\mathcal C,\bar\chi,\chi^{ij}\}$, and the tertiary constraint
$\theta$ therefore have the bracket pattern shown in
Table~\ref{tab:PBdegen}. Here and in the following tables, ``0'' denotes a
bracket that vanishes weakly as an integral kernel, whereas ``$X$'' denotes an
entry not forced to vanish. A diagonal ``$X$'' is compatible with
antisymmetry when the bracket contains derivatives of delta functions.
\begin{center}
	\begin{table}[h]
		\begin{centering}
			\begin{tabular}{c|cccc|cccc|c}
				$[\bullet,\bullet]$ & $\pi_i(\vec y)$ & $\pi_N(\vec y)$ & $\bar p_F(\vec y)$ & $p^{kl}(\vec y)$ & $\mathcal C_i(\vec y)$ & $\mathcal C(\vec y)$ & $\bar\chi(\vec y)$ & $\chi^{ij}(\vec y)$ & $\theta(\vec y)$\tabularnewline
				\hline
				$\pi_i(\vec x)$ & 0 & 0 & 0 & 0 & 0 & 0 & 0 & 0 & 0\tabularnewline
				$\pi_N(\vec x)$ & 0 & 0 & 0 & 0 & 0 & $X$ & 0 & $X$ & $X$\tabularnewline
				$\bar p_F(\vec x)$ & 0 & 0 & 0 & 0 & 0 & 0 & 0 & 0 & $X$\tabularnewline
				$p^{kl}(\vec x)$ & 0 & 0 & 0 & 0 & 0 & $X$ & 0 & $X$ & $X$\tabularnewline
				\hline
				$\mathcal C_i(\vec x)$ & 0 & 0 & 0 & 0 & 0 & 0 & 0 & 0 & 0\tabularnewline
				$\mathcal C(\vec x)$ & 0 & $X$ & 0 & $X$ & 0 & $X$ & $X$ & $X$ & $X$\tabularnewline
				$\bar\chi(\vec x)$ & 0 & 0 & 0 & 0 & 0 & $X$ & $X$ & $X$ & $X$\tabularnewline
				$\chi^{ij}(\vec x)$ & 0 & $X$ & 0 & $X$ & 0 & $X$ & $X$ & $X$ & $X$\tabularnewline
				\hline
				$\theta(\vec x)$ & 0 & $X$ & $X$ & $X$ & 0 & $X$ & $X$ & $X$ & $X$\tabularnewline
			\end{tabular}
			\par
		\end{centering}
		\caption{Poisson brackets among the zero-mode-adapted primary, secondary, and tertiary constraints after imposing the degeneracy condition.}
		\label{tab:PBdegen}
	\end{table}
	\par
\end{center}

Assuming that the remaining 17-component Dirac operator has maximal
functional rank, the six constraints $\pi_i$ and $\mathcal C_i$ are
first-class and the other 17 are second-class. The formal Dirac count is then
\begin{align}
	\#_{\mathrm{DOF}}
	&=\frac{1}{2}\left(
	2\times\#_{\mathrm{var}}
	-2\times\#_{1\mathrm{st}}
	-\#_{2\mathrm{nd}}\right)\nonumber\\
	&=\frac{1}{2}\left(2\times19-2\times6-17\right)
	=4.5.
\end{align}
This half-integer is not the count of an additional healthy particle
polarization. It signals that degeneracy alone has removed only one direction
of the unwanted canonical pair. In a field theory, an antisymmetric
non-ultralocal Dirac operator can have odd functional rank because derivatives
of delta functions invalidate the finite-dimensional even-rank argument. An
analogous formal half-integer count and the need for a further condition occur
in spatially covariant gravity with a lapse velocity
\citep{Gao:2018znj}. The remaining half mode must therefore be removed by an
additional condition.

\subsection{The consistency condition}

Under the degeneracy condition and apart from the spatial-diffeomorphism sector $\pi_i$, $\bar p_F$ is the unique primary combination that commutes weakly with every primary and secondary constraint. Upon adding
$\theta$, the only new bracket that can lift this zero-mode direction is
$[\bar p_F,\theta]$. Requiring this projected bracket to vanish is therefore
not an arbitrary choice: it is the rank condition that extends the zero mode
of the primary-secondary block to the enlarged constraint algebra. It allows
the remaining half of the unwanted canonical pair to be removed either by an
additional first-class direction or by a quaternary constraint.

For independent smearing functions $\tilde V$ and $V$, consider
\begin{equation}
	\left[
	\int\mathrm{d}^{3}x\,\tilde V(\vec{x})\bar p_F(\vec{x}),
	\int\mathrm{d}^{3}y\,V(\vec{y})\theta(\vec{y})
	\right]
	\approx
	\int\mathrm{d}^{3}x\mathrm{d}^{3}y\,
	\tilde V(\vec{x})V(\vec{y})
	[\bar p_F(\vec{x}),\theta(\vec{y})].
	\label{eq:PBpFtht}
\end{equation}
Using Eqs.~\eqref{eq:redef_barchi} and
\eqref{eq:redefinition equation of p_F}, and dropping terms proportional to
existing constraints, the left-hand side reduces to the two essential terms
\begin{equation}
	\mathrm{L.H.S.}\approx
	\int\mathrm{d}^{3}x\mathrm{d}^{3}y\,
	\tilde V_{\alpha}(\vec{x})
	\Bigl(
	[\varphi^{\alpha}(\vec{x}),V_{\beta}(\vec{y})]
	\mathfrak D_0\chi^{\beta}(\vec{y})
	+V_{\beta}(\vec{y})
	[\varphi^{\alpha}(\vec{x}),
	\mathfrak D_0\chi^{\beta}(\vec{y})]
	\Bigr).
	\label{eq:lhsPBpFtht}
\end{equation} 
The operator
$\mathfrak{D}_{0}$ is a derivation of the Poisson bracket and therefore gives
\begin{equation}
	[\varphi^{\alpha}(\vec{x}),
	\mathfrak{D}_{0}\chi^{\beta}(\vec{y})]
	=
	-\mathfrak{D}_{0}\mathcal P^{\beta\alpha}(\vec{y},\vec{x})
	+[\chi^{\beta}(\vec{y}),\chi^{\alpha}(\vec{x})].
\end{equation}
Applying this identity to Eq.~\eqref{eq:lhsPBpFtht} yields 
\begin{align}
	\mathrm{L.H.S.}\approx{}&
	\int\mathrm{d}^{3}y\,
	Z_{\beta}(\vec{y})\mathfrak D_0\chi^{\beta}(\vec{y})
	+\int\mathrm{d}^{3}x\mathrm{d}^{3}y\,
	\tilde V_{\alpha}(\vec{x})V_{\beta}(\vec{y})
	[\chi^{\beta}(\vec{y}),\chi^{\alpha}(\vec{x})]\nonumber\\
	& -\int\mathrm{d}^{3}x\mathrm{d}^{3}y\,
	\tilde V_{\alpha}(\vec{x})V_{\beta}(\vec{y})
	\mathfrak D_0\mathcal P^{\beta\alpha}(\vec{y},\vec{x}),
	\label{eq:PBpFtht2}
\end{align}
where
\begin{equation}
	Z_{\beta}(\vec{y})\coloneqq
	\int\mathrm{d}^{3}x\,
	\tilde V_{\alpha}(\vec{x})
	[\varphi^{\alpha}(\vec{x}),V_{\beta}(\vec{y})].
\end{equation}

In order to proceed, note that since 
$V_{\alpha}$ in \eqref{eq:null vector} is a weak left-zero-mode, there must be
coefficients $\mathcal A^{\beta}{}_{\gamma}$ and
$\mathcal B^{\beta}{}_{\gamma}$ such that,
\begin{equation}
	\int\mathrm{d}^{3}x\,
	V_{\alpha}(\vec{x})
	\mathcal P^{\alpha\beta}(\vec{x},\vec{y})\equiv
	\int\mathrm{d}^{3}z\,
	\left[
	\mathcal A^{\beta}{}_{\gamma}(\vec{y},\vec{z})
	\chi^{\gamma}(\vec{z})
	+\mathcal B^{\beta}{}_{\gamma}(\vec{y},\vec{z})
	\varphi^{\gamma}(\vec{z})
	\right] \approx 0.
	\label{eq:left-null-remainder}
\end{equation}
Taking $\mathfrak D_0$ of \eqref{eq:left-null-remainder} and contracting with
the right-zero-mode, we obtain
\begin{equation}
	\int\mathrm{d}^{3}x\mathrm{d}^{3}y\,
	\tilde V_{\alpha}(\vec{x})V_{\beta}(\vec{y})
	\mathfrak D_0\mathcal P^{\beta\alpha}(\vec{y},\vec{x})
	\approx
	\int\mathrm{d}^{3}x\mathrm{d}^{3}y\,
	\tilde V_{\alpha}(\vec{x})
	\mathcal A^{\alpha}{}_{\beta}(\vec{x},\vec{y})
	\mathfrak D_0\chi^{\beta}(\vec{y}),
\end{equation}
where we used $\mathfrak D_0\varphi^\alpha=\chi^\alpha$.
In deriving the above, terms on which $\mathfrak D_0$ acts on $V_\beta$ are vanishing weakly after contracting with the right-zero-mode. Therefore, \eqref{eq:PBpFtht2} now becomes
\begin{equation}
	\mathrm{L.H.S.}\approx
	\int\mathrm{d}^{3}y\,
	\bar{Z}_{\beta}(\vec{y})\mathfrak D_0\chi^{\beta}(\vec{y})
	+\int\mathrm{d}^{3}x\mathrm{d}^{3}y\,
	\tilde V_{\alpha}(\vec{x})V_{\beta}(\vec{y})
	[\chi^{\beta}(\vec{y}),\chi^{\alpha}(\vec{x})],
	\label{eq:PBpFtht3}
\end{equation}
where we define
\begin{equation}
	\bar{Z}_{\beta}(\vec{y})\coloneqq
	Z_{\beta}(\vec{y})
	-\int\mathrm{d}^{3} x\,
	\tilde V_{\alpha}(\vec{x})
	\mathcal A^{\alpha}{}_{\beta}(\vec{x},\vec{y}).
\end{equation}
We now show that $\bar{Z}_\beta$ is a weak left-zero-mode. By using the
Leibniz rule, the Jacobi identity, and both weak zero-mode relations, we find
\begin{align}
	\int\mathrm{d}^{3}y\,
	Z_{\beta}(\vec{y})\mathcal P^{\beta\sigma}(\vec{y},\vec{z})
	&=\int\mathrm{d}^{3}x\mathrm{d}^{3}y\,
	\tilde V_{\alpha}(\vec{x})
	\Bigl\{
	[\varphi^{\alpha}(\vec{x}),
	V_{\beta}(\vec{y})\mathcal P^{\beta\sigma}(\vec{y},\vec{z})]
	-V_{\beta}(\vec{y})
	[\varphi^{\alpha}(\vec{x}),
	\mathcal P^{\beta\sigma}(\vec{y},\vec{z})]
	\Bigr\}\nonumber\\
	&=\int\mathrm{d}^{3}x\mathrm{d}^{3}y\,
	\tilde V_{\alpha}(\vec{x})
	\Bigl\{
	[\varphi^{\alpha}(\vec{x}),
	V_{\beta}(\vec{y})\mathcal P^{\beta\sigma}(\vec{y},\vec{z})]
	-V_{\beta}(\vec{y})
	[\varphi^{\sigma}(\vec{z}),
	\mathcal P^{\beta\alpha}(\vec{y},\vec{x})]
	\Bigr\}\nonumber\\
	&\approx\int\mathrm{d}^{3}x\mathrm{d}^{3}y\,
	\tilde V_{\alpha}(\vec{x})
	\mathcal A^{\alpha}{}_{\beta}(\vec{x},\vec{y})
	\mathcal P^{\beta\sigma}(\vec{y},\vec{z}).
	\label{eq:intZcalP}
\end{align}
It follows immediately that
\begin{equation}
	\int\mathrm{d}^{3}y\,
	\bar{Z}_{\beta}(\vec{y})
	\mathcal P^{\beta\sigma}(\vec{y},\vec{z})\approx0.
\end{equation}

Since the local kernel has one generator, there is a smearing function
$g(\vec y)$, depending bilinearly on $V$ and $\tilde V$, such that 
\begin{equation}
	\bar{Z}_{\beta}(\vec{y})
	\approx
	\left(
	U[g](\vec{y}),\,g(\vec{y}),\,V_{ij}[g](\vec{y})
	\right),
	\label{eq:Z-functional}
\end{equation}
where
\begin{align}
	U[g](\vec{y})&\equiv
	\int\mathrm{d}^{3}z\,
	g(\vec{z})\mathcal U(\vec{z},\vec{y}),\\
	V_{ij}[g](\vec{y})&\equiv
	\int\mathrm{d}^{3}z\,
	g(\vec{z})\mathcal V_{ij}(\vec{z},\vec{y}).
\end{align}
It follows that the first term in Eq.~\eqref{eq:PBpFtht3} vanishes weakly (on the enlarged constraint surface including $\theta \approx 0$), 
\begin{align}
	\int\mathrm{d}^{3}y\,
	\bar{Z}_{\beta}(\vec{y})\mathfrak D_0\chi^{\beta}(\vec{y})
	&\approx\mathfrak D_0
	\int\mathrm{d}^{3}y\,
	g(\vec{y})\bar\chi(\vec{y})\nonumber\\
	&\approx\int\mathrm{d}^{3}y\,
	g(\vec{y})\theta(\vec{y})\approx0.
\end{align}
Again, we used that terms in which $\mathfrak{D}_{0}$ acts on $g$ are proportional to $\bar\chi$ and therefore vanish weakly.
Therefore,
\begin{equation}
	\left[
	\int\mathrm{d}^{3}x\,\tilde V(\vec{x})\bar p_F(\vec{x}),
	\int\mathrm{d}^{3}y\,V(\vec{y})\theta(\vec{y})
	\right]
	\approx
	\int\mathrm{d}^{3}x\mathrm{d}^{3}y\,
	\tilde V_{\alpha}(\vec{x})V_{\beta}(\vec{y})
	[\chi^{\beta}(\vec{y}),\chi^{\alpha}(\vec{x})].
	\label{eq:projected-secondary-bracket}
\end{equation}

Using Eqs.~\eqref{eq:V_ij explicitlt}, \eqref{eq:U explicitly}, and
\eqref{eq:V_ij~U~}, the projected bracket can be written as
\begin{equation}
	\int\mathrm{d}^{3}x\mathrm{d}^{3}y\,
	\tilde V_{\alpha}(\vec{x})V_{\beta}(\vec{y})
	[\chi^{\beta}(\vec{y}),\chi^{\alpha}(\vec{x})]
	\approx
	\int\mathrm{d}^{3}x\mathrm{d}^{3}y\,
	\tilde V(\vec{x})V(\vec{y})\mathcal F(\vec{x},\vec{y}).
	\label{eq:-2}
\end{equation}
After carrying out the delta-function integrations, the kernel is
\begin{empheq}[box=\fbox]{align}
	\mathcal F(\vec{x},\vec{y})\coloneqq{}&
	\int\mathrm{d}^{3}z\mathrm{d}^{3}z'\,
	\tilde{\mathcal U}(\vec{x},\vec{z})
	\mathcal U(\vec{y},\vec{z}')
	[\mathcal C(\vec{z}'),\mathcal C(\vec{z})]\nonumber\\
	&+\int\mathrm{d}^{3}z\,
	\tilde{\mathcal U}(\vec{x},\vec{z})
	[\chi(\vec{y}),\mathcal C(\vec{z})]
	+\int\mathrm{d}^{3}z\mathrm{d}^{3}z'\,
	\tilde{\mathcal U}(\vec{x},\vec{z})
	\mathcal V_{ij}(\vec{y},\vec{z}')
	[\chi^{ij}(\vec{z}'),\mathcal C(\vec{z})]\nonumber\\
	&+\int\mathrm{d}^{3}z'\,
	\mathcal U(\vec{y},\vec{z}')
	[\mathcal C(\vec{z}'),\chi(\vec{x})]
	+[\chi(\vec{y}),\chi(\vec{x})]
	+\int\mathrm{d}^{3}z'\,
	\mathcal V_{ij}(\vec{y},\vec{z}')
	[\chi^{ij}(\vec{z}'),\chi(\vec{x})]\nonumber\\
	&+\int\mathrm{d}^{3}z\mathrm{d}^{3}z'\,
	\tilde{\mathcal V}_{kl}(\vec{x},\vec{z})
	\mathcal U(\vec{y},\vec{z}')
	[\mathcal C(\vec{z}'),\chi^{kl}(\vec{z})]
	+\int\mathrm{d}^{3}z\,
	\tilde{\mathcal V}_{kl}(\vec{x},\vec{z})
	[\chi(\vec{y}),\chi^{kl}(\vec{z})]\nonumber\\
	&+\int\mathrm{d}^{3}z\mathrm{d}^{3}z'\,
	\tilde{\mathcal V}_{kl}(\vec{x},\vec{z})
	\mathcal V_{ij}(\vec{y},\vec{z}')
	[\chi^{ij}(\vec{z}'),\chi^{kl}(\vec{z})].\label{eq:conscalFxpl}
\end{empheq}
Comparison with Eq.~\eqref{eq:PBpFtht} gives
\begin{equation}
	[\bar p_F(\vec{x}),\theta(\vec{y})]
	\approx\mathcal F(\vec{x},\vec{y}).
	\label{eq:consistency condition_0}
\end{equation}
We impose the additional consistency condition
\begin{equation}
	\boxed{\mathcal F(\vec{x},\vec{y})\approx0.}
	\label{eq:consistency condition}
\end{equation}
Here the weak equality is evaluated on the enlarged constraint surface that includes $\theta\approx0$.  After the momenta are eliminated
with the existing constraints, it becomes a functional identity restricting
the Lagrangian. Equation~\eqref{eq:consistency condition} is important because
it ensures that the zero mode of the primary-secondary bracket remains a zero
mode after the tertiary constraint is included. The name refers to this
compatibility of the successive constraint-algebra blocks, not to the generic
Dirac-Bergmann requirement of preservation in time.

We next examine the preservation of $\theta$. Once
Eqs.~\eqref{eq:degeneracy condition} and
\eqref{eq:consistency condition} hold, its preservation equation is
\begin{equation}
	\mathfrak D_0\theta(\vec{x})
	+\int\mathrm{d}^{3}y\,
	[\theta(\vec{x}),\pi_N(\vec{y})]\lambda(\vec{y})
	+\int\mathrm{d}^{3}y\,
	[\theta(\vec{x}),p^{kl}(\vec{y})]v_{kl}(\vec{y})
	\approx0.
	\label{eq:evolution of =00003D00005Ctheta}
\end{equation}
The multiplier of $\bar p_F$ is absent by
Eq.~\eqref{eq:consistency condition}. Because the corank of $\mathcal P$ is
one, the complementary $7\times7$ operator is invertible and the preservation
equations for $\mathcal C$ and $\chi^{ij}$,
\begin{align}
	\mathfrak D_0\mathcal C(\vec{x})
	&+\int\mathrm{d}^{3}y\,
	[\mathcal C(\vec{x}),\pi_N(\vec{y})]\lambda(\vec{y})
	+\int\mathrm{d}^{3}y\,
	[\mathcal C(\vec{x}),p^{kl}(\vec{y})]v_{kl}(\vec{y})\approx0,\\
	\mathfrak D_0\chi^{ij}(\vec{x})
	&+\int\mathrm{d}^{3}y\,
	[\chi^{ij}(\vec{x}),\pi_N(\vec{y})]\lambda(\vec{y})
	+\int\mathrm{d}^{3}y\,
	[\chi^{ij}(\vec{x}),p^{kl}(\vec{y})]v_{kl}(\vec{y})\approx0,
\end{align}
determine $\lambda=\lambda_*$ and $v_{ij}=v^*_{ij}$ under the assumed
regularity conditions. Substitution into
Eq.~\eqref{eq:evolution of =00003D00005Ctheta} gives two possibilities:
\begin{itemize}
	\item In Case I, the resulting expression is a combination of existing
	constraints and vanishes weakly. The preservation of $\theta$ then produces
	no further constraint.
	\item In Case II, the resulting expression is independent of the existing
	constraints and defines the quaternary constraint
	\begin{equation}
		\boxed{\omega(\vec{x})\coloneqq
		\mathfrak D_0\theta(\vec{x})
		+\int\mathrm{d}^{3}y\,
		[\theta(\vec{x}),\pi_N(\vec{y})]\lambda_*(\vec{y})
		+\int\mathrm{d}^{3}y\,
		[\theta(\vec{x}),p^{kl}(\vec{y})]v^*_{kl}(\vec{y})
		\approx0.}
		\label{eq:quaternary constraint}
	\end{equation}
\end{itemize}
Both cases propagate four physical degrees of freedom, as shown next.

\subsection{The physical degrees of freedom}

We now count the physical degrees of freedom after imposing both structural
conditions. The tables display only vanishings established by the preceding
zero-mode analysis and spatial covariance.
The counts additionally assume that each indicated second-class suboperator
has maximal functional rank.

In Case I, the brackets are summarized in Table~\ref{tab:PBcase1}.
\begin{table}[H]
	\begin{centering}
		\begin{tabular}{c|cccc|cccc|c}
			$[\bullet,\bullet]$ & $\pi_i(\vec y)$ & $\pi_N(\vec y)$ & $\bar p_F(\vec y)$ & $p^{kl}(\vec y)$ & $\mathcal C_i(\vec y)$ & $\mathcal C(\vec y)$ & $\bar\chi(\vec y)$ & $\chi^{ij}(\vec y)$ & $\theta(\vec y)$\tabularnewline
			\hline
			$\pi_i(\vec x)$ & 0 & 0 & 0 & 0 & 0 & 0 & 0 & 0 & 0\tabularnewline
			$\pi_N(\vec x)$ & 0 & 0 & 0 & 0 & 0 & $X$ & 0 & $X$ & $X$\tabularnewline
			$\bar p_F(\vec x)$ & 0 & 0 & 0 & 0 & 0 & 0 & 0 & 0 & 0\tabularnewline
			$p^{kl}(\vec x)$ & 0 & 0 & 0 & 0 & 0 & $X$ & 0 & $X$ & $X$\tabularnewline
			\hline
			$\mathcal C_i(\vec x)$ & 0 & 0 & 0 & 0 & 0 & 0 & 0 & 0 & 0\tabularnewline
			$\mathcal C(\vec x)$ & 0 & $X$ & 0 & $X$ & 0 & $X$ & $X$ & $X$ & $X$\tabularnewline
			$\bar\chi(\vec x)$ & 0 & 0 & 0 & 0 & 0 & $X$ & $X$ & $X$ & $X$\tabularnewline
			$\chi^{ij}(\vec x)$ & 0 & $X$ & 0 & $X$ & 0 & $X$ & $X$ & $X$ & $X$\tabularnewline
			\hline
			$\theta(\vec x)$ & 0 & $X$ & 0 & $X$ & 0 & $X$ & $X$ & $X$ & $X$\tabularnewline
		\end{tabular}
		\par
	\end{centering}
	\caption{Poisson brackets among all constraints in Case I.}
	\label{tab:PBcase1}
\end{table}
The preservation of $\bar p_F$ gives
$\mathfrak D_0\bar p_F\approx\bar\chi$ up to primary constraints, so it is
already satisfied on the constraint surface. Moreover,
Eq.~\eqref{eq:consistency condition} makes its bracket with $\theta$ vanish.
Thus $\pi_i$, $\mathcal C_i$, and $\bar p_F$ are seven first-class
constraints, while $\pi_N$, $p^{ij}$, $\mathcal C$, $\bar\chi$, $\chi^{ij}$,
and $\theta$ are 16 second-class constraints. Hence
\begin{align}
	\#_{\mathrm{DOF}}
	&=\frac{1}{2}\left(2\times19-2\times7-16\right)
	=4.
\end{align}

In Case II, the brackets are summarized in Table~\ref{tab:PBcase2}.
\begin{table}[H]
	\begin{centering}
		\begin{tabular}{c|cccc|cccc|cc}
			$[\bullet,\bullet]$ & $\pi_i(\vec y)$ & $\pi_N(\vec y)$ & $\bar p_F(\vec y)$ & $p^{kl}(\vec y)$ & $\mathcal C_i(\vec y)$ & $\mathcal C(\vec y)$ & $\bar\chi(\vec y)$ & $\chi^{ij}(\vec y)$ & $\theta(\vec y)$ & $\omega(\vec y)$\tabularnewline
			\hline
			$\pi_i(\vec x)$ & 0 & 0 & 0 & 0 & 0 & 0 & 0 & 0 & 0 & 0\tabularnewline
			$\pi_N(\vec x)$ & 0 & 0 & 0 & 0 & 0 & $X$ & 0 & $X$ & $X$ & $X$\tabularnewline
			$\bar p_F(\vec x)$ & 0 & 0 & 0 & 0 & 0 & 0 & 0 & 0 & 0 & $X$\tabularnewline
			$p^{kl}(\vec x)$ & 0 & 0 & 0 & 0 & 0 & $X$ & 0 & $X$ & $X$ & $X$\tabularnewline
			\hline
			$\mathcal C_i(\vec x)$ & 0 & 0 & 0 & 0 & 0 & 0 & 0 & 0 & 0 & 0\tabularnewline
			$\mathcal C(\vec x)$ & 0 & $X$ & 0 & $X$ & 0 & $X$ & $X$ & $X$ & $X$ & $X$\tabularnewline
			$\bar\chi(\vec x)$ & 0 & 0 & 0 & 0 & 0 & $X$ & $X$ & $X$ & $X$ & $X$\tabularnewline
			$\chi^{ij}(\vec x)$ & 0 & $X$ & 0 & $X$ & 0 & $X$ & $X$ & $X$ & $X$ & $X$\tabularnewline
			\hline
			$\theta(\vec x)$ & 0 & $X$ & 0 & $X$ & 0 & $X$ & $X$ & $X$ & $X$ & $X$\tabularnewline
			$\omega(\vec x)$ & 0 & $X$ & $X$ & $X$ & 0 & $X$ & $X$ & $X$ & $X$ & $X$\tabularnewline
		\end{tabular}
		\par
	\end{centering}
	\caption{Poisson brackets among all constraints in Case II.}
	\label{tab:PBcase2}
\end{table}
There are now 24 constraints: the six constraints $\pi_i$ and $\mathcal C_i$
are first-class, and the remaining 18 are second-class. The degree-of-freedom
count is
\begin{align}
	\#_{\mathrm{DOF}}
	&=\frac{1}{2}\left(2\times19-2\times6-18\right)
	=4.
\end{align}
After $\lambda_*$ and $v^*_{ij}$ have been substituted, preservation of
$\omega$ contains the one still-undetermined multiplier $\bar v$ of
$\bar p_F$ through
$\int\mathrm d^3y\,[\omega(\vec x),\bar p_F(\vec y)]\bar v(\vec y)$.
If this residual operator is invertible on the chosen function space, the
equation fixes $\bar v$ and the chain terminates. If it has a kernel, further
constraints or first-class directions can occur, which is a different branch
from Case II as counted here.

The six first-class constraints $\pi_i$ and $\mathcal C_i$ in both cases
generate the residual spatial diffeomorphisms. In Case I, the additional
first-class direction $\bar p_F$ indicates, under the standard regularity
assumptions, one further canonical gauge redundancy. Its infinitesimal action
on the auxiliary configuration variables is along the right zero mode in
Eq.~\eqref{eq:redefine p_F} and therefore mixes $F$, $N$, and $B_{ij}$. This is
an enhancement beyond spatial diffeomorphisms, but it does not by itself imply
restoration of temporal diffeomorphisms or full spacetime covariance.

We close with two qualifications concerning the scope of the result. The
degeneracy of the full constraint operator can be realized in several branches.
Here we assumed from the outset that
$\delta^2S/\delta N(\vec x)\delta N(\vec y)$ and the effective tensor block
$\mathcal J^{ij,kl}(\vec x,\vec y)$ are invertible, and that
$\mathcal P^{\alpha\beta}$ has one local zero-mode direction. The degeneracy
and consistency conditions derived above therefore provide one mechanism for
obtaining four degrees of freedom under these assumptions, rather than a
classification of every degenerate theory contained in Eq.~\eqref{eq:S0}.

The bi-Galileon theory displayed in Appendix~\ref{app:bigalug} illustrates this
distinction. After integrations by parts, its unitary-gauge action belongs to
the broad class \eqref{eq:S0}, but this does not imply that it lies in the
specific branch analyzed here. In that representation $F=\pounds_{\bm u}^2\psi$
enters linearly and has no kinetic mixing with $B_{ij}$, so that
$\delta^2S/\delta F^2=0$ and
$\delta^2S/(\delta F\,\delta B_{ij})=0$. Depending on the remaining lapse
couplings, the invertibility or kernel assumptions used above may fail. A more
natural reduction for this structure starts from an invertible $B_{ij}$ block
and then analyzes the residual $(N,F)$ operator. That alternative branch lies
outside the present analysis. Related multi-field degeneracy structures are
studied in Ref.~\cite{BouzariNezhad:2026zsj}.

\section{Multi-field disformal transformation and its implication} \label{sec:dftrans}

Field redefinitions provide a useful way to relate different formulations and to generate new scalar-tensor theories. It is thus interesting to examine whether such a transformation preserves the class of spatially covariant scalar field actions in Eq.~(\ref{eq:S0}), in particular its exclusion of the velocity of the lapse.

\subsection{Multi-field disformal transformation}

The single-field disformal transformation introduced in \cite{Bekenstein:1992pj} generalizes a conformal rescaling by adding a term along the scalar gradient,
\begin{equation}
	\widetilde{g}_{\mu\nu}=\varOmega(\phi,X)g_{\mu\nu}+\varGamma(\phi,X)\nabla_{\mu}\phi\nabla_{\nu}\phi,
	\qquad
	X\coloneqq-\frac{1}{2}g^{\mu\nu}\nabla_{\mu}\phi\nabla_{\nu}\phi.
\end{equation}
Consider now $N_{\mathrm{s}}$ scalar fields $\phi^I$ with $I=1,\ldots,N_{\mathrm{s}}$.
A multi-field extension of the above disformal transformation was proposed in \cite{Watanabe:2015uqa}:
\begin{equation}
	\widetilde{g}_{\mu\nu}=\varOmega\left(\phi^{K},X^{KL}\right)g_{\mu\nu}+\varGamma_{IJ}\left(\phi^{K},X^{KL}\right)\nabla_{\mu}\phi^{I}\nabla_{\nu}\phi^{J},\label{eq:disformal trans}
\end{equation}
where $X^{IJ}\coloneqq-\frac{1}{2}g^{\mu\nu}\nabla_{\mu}\phi^{I}\nabla_{\nu}\phi^{J}$.

The inverse metric can be written as
\begin{equation}
	\widetilde{g}^{\mu\nu}=\varOmega^{-1}\left(g^{\mu\nu}-Z_{IJ}\nabla^{\mu}\phi^{I}\nabla^{\nu}\phi^{J}\right).
\end{equation}
Requiring $\widetilde{g}_{\mu\rho}\widetilde{g}^{\rho\nu}=\delta_{\mu}^{\nu}$ determines $Z_{IJ}$ through
\begin{equation}
	\varOmega Z_{IJ}-2\varGamma_{IA}X^{AB}Z_{BJ}=\varGamma_{IJ}. \label{dtZcond}
\end{equation}
Thus, if $\varOmega\neq 0$ and the field-space matrix
$M_{I}{}^{J}\coloneqq\varOmega\delta_{I}^{J}-2\varGamma_{IA}X^{AJ}$ is nonsingular, Eq.~(\ref{dtZcond}) has the unique solution $Z=M^{-1}\varGamma$.

The transformed kinetic matrix is
\begin{align}
	\widetilde{X}^{IJ} & \coloneqq -\frac{1}{2}\widetilde{g}^{\mu\nu}\nabla_{\mu}\phi^{I}\nabla_{\nu}\phi^{J}
	=\varOmega^{-1}\left(X^{IJ}+2Z_{AB}X^{AI}X^{BJ}\right).
\end{align}

For the metric determinant, the matrix determinant lemma gives
\begin{align}
	\widetilde{g}
	&=\varOmega^{4}g\,\det
	\left(\delta^{I}_{J}-2\varOmega^{-1}\varGamma_{JK}X^{KI}\right).
\end{align}
We now specialize to two scalar fields. In this case,
\begin{equation}
	\widetilde{g}=\varOmega^{2}g\,\det\left(\varOmega\delta^{I}_{J}-2\varGamma_{JK}X^{KI}\right),
\end{equation}
and, choosing the signature-preserving branch with $\varOmega>0$ and a positive field-space determinant,
\begin{equation}
	\sqrt{-\widetilde{g}}=\varOmega\sqrt{-g}\,
	\sqrt{\det\left(\varOmega\delta^{I}_{J}-2\varGamma_{JK}X^{KI}\right)}.\label{eq:trans fo =00005Csqrtg}
\end{equation}
Using $\det(\varOmega\mathbf{I}-\mathbf{A})=\varOmega^{2}-\varOmega\operatorname{Tr}(\mathbf{A})+\det(\mathbf{A})$ for a $2\times2$ matrix, this becomes
\begin{equation}
	\sqrt{-\widetilde{g}}=\varOmega\sqrt{-g}
	\sqrt{\varOmega^{2}-2\varOmega\varGamma_{IJ}X^{IJ}
		+4\det(\varGamma_{IJ})\det(X^{IJ})}.
\end{equation}

Nonsingularity of $\widetilde g_{\mu\nu}$ is distinct from local invertibility of the map $g_{\mu\nu}\mapsto\widetilde g_{\mu\nu}$ when the transformation functions depend on $X^{IJ}$. The latter is controlled by the Jacobian \citep{Firouzjahi:2018xob}
\begin{equation}
	J^{\alpha\beta}_{\mu\nu}\coloneqq\frac{\partial\widetilde{g}_{\mu\nu}}{\partial g_{\alpha\beta}}
	=\varOmega\delta^{(\alpha}_{\mu}\delta^{\beta)}_{\nu}
	+\frac{1}{2}\left(g_{\mu\nu}\frac{\partial\varOmega}{\partial X^{KL}}
	+\nabla_{\mu}\phi^{I}\nabla_{\nu}\phi^{J}\frac{\partial\varGamma_{IJ}}{\partial X^{KL}}\right)
	\nabla^{(\alpha}\phi^{K}\nabla^{\beta)}\phi^{L}.
\end{equation}
Let $v_{\mu\nu}$ be an eigentensor,
\begin{equation}
	J^{\alpha\beta}_{\mu\nu}v_{\alpha\beta}=\lambda v_{\mu\nu},\label{eq:eigen of Jacobi}
\end{equation}
and define the symmetric field-space tensor
\begin{equation}
	C^{KL}\coloneqq \nabla^{\alpha}\phi^{K}\nabla^{\beta}\phi^{L}v_{\alpha\beta}.
\end{equation}
Equation~(\ref{eq:eigen of Jacobi}) is then equivalent to
\begin{equation}
	(\lambda-\varOmega)v_{\mu\nu}=\frac{1}{2}\left[g_{\mu\nu}\frac{\partial\varOmega}{\partial X^{KL}}
	+\nabla_{\mu}\phi^{I}\nabla_{\nu}\phi^{J}\frac{\partial\varGamma_{IJ}}{\partial X^{KL}}\right]C^{KL}.
\end{equation}
For two generically independent scalar gradients, $C^{KL}=0$ imposes three independent linear conditions on the ten-dimensional space of symmetric spacetime tensors. The corresponding seven-dimensional kernel therefore has eigenvalue $\lambda=\varOmega$. On nongeneric field configurations the kernel can be larger, but the determinant formula below remains valid. Contracting the eigenvalue equation with $\nabla^{\mu}\phi^{M}\nabla^{\nu}\phi^{N}$ gives 
\begin{equation}
	(\lambda-\varOmega)C^{MN}=\left[-X^{MN}\frac{\partial\varOmega}{\partial X^{KL}}
	+2X^{IM}X^{JN}\frac{\partial\varGamma_{IJ}}{\partial X^{KL}}\right]C^{KL}.
\end{equation}
Consequently,
\begin{equation}
	\det(J^{\alpha\beta}_{\mu\nu})=\varOmega^{7}\det_{3\times3}
	\left[\varOmega\delta^{(M}_{K}\delta^{N)}_{L}
	-X^{MN}\frac{\partial\varOmega}{\partial X^{KL}}
	+2X^{MI}X^{NJ}\frac{\partial\varGamma_{IJ}}{\partial X^{KL}}\right].
\end{equation}
Subject also to the metric nonsingularity and signature conditions stated above, the inverse-function theorem therefore gives the local invertibility conditions
\begin{equation}
	\varOmega\neq0,
	\qquad
	\det_{3\times3}\left[\varOmega\delta^{(M}_{K}\delta^{N)}_{L}
	-X^{MN}\frac{\partial\varOmega}{\partial X^{KL}}
	+2X^{MI}X^{NJ}\frac{\partial\varGamma_{IJ}}{\partial X^{KL}}\right]\neq0.
\end{equation}

The inverse-metric relation and the field-space determinant above hold for arbitrary $N_{\mathrm{s}}$, whereas the expanded determinant, the $7+3$ decomposition, and the $3\times3$ Jacobian criterion are specific to two fields. More generally, if the scalar gradients span an $r$-dimensional subspace of the cotangent space, with $r\leq\min(N_{\mathrm{s}},4)$, the nontrivial part of the Jacobian acts on at most $r(r+1)/2$ independent symmetric gradient combinations. For generic independent gradients this gives a $6\times6$ reduced problem for three fields and a $10\times10$ problem for four fields. Additional scalar gradients will not increase this number beyond ten. The explicit analysis below therefore focuses on the bi-scalar case, for which the reduction is most useful.

\subsection{Bi-field disformal transformation in the unitary gauge}

Let $\phi^{1}=\phi$ and $\phi^{2}=\psi$. In the unitary gauge $\phi=t$, the kinetic terms are
\begin{equation}
	X^{11}=\frac{1}{2N^{2}},\qquad
	X^{12}=\frac{1}{2N}\pounds_{\bm{u}}\psi,\qquad
	X^{22}=\frac{1}{2}\left(\pounds_{\bm{u}}\psi\right)^{2}
	-\frac{1}{2}\mathrm{D}_{i}\psi\mathrm{D}^{i}\psi\eqqcolon Y.
\end{equation}
Thus $\varOmega$ and $\varGamma_{IJ}$ can be regarded as functions of $t$, $N$, $\psi$, $\pounds_{\bm{u}}\psi$, and $(\mathrm D\psi)^2\coloneqq\mathrm D_i\psi\mathrm D^i\psi$ in this gauge.

The ADM variables transform as
\begin{align}
	\widetilde{N}&=N\varPhi\left(t,N,\psi,\pounds_{\bm{u}}\psi,(\mathrm{D}\psi)^2\right),\label{eq:trans of N}
	\\
	\widetilde{N}_{i}&=\varOmega N_{i}+\left(\varGamma_{12}+\varGamma_{22}\partial_{t}\psi\right)\mathrm{D}_{i}\psi,\label{eq:trans of N_i}
	\\
	\widetilde{h}_{ij}&=\varOmega h_{ij}+\varGamma_{22}\mathrm{D}_{i}\psi\mathrm{D}_{j}\psi.\label{eq:trans of h_ij}
\end{align}
Here
$\partial_t\psi=N\pounds_{\bm u}\psi+N^i\mathrm D_i\psi$ is the derivative with respect to the time coordinate. The lapse factor is
\begin{equation}
	\varPhi\coloneqq\frac{\sqrt{-\widetilde{g}}}{\sqrt{-g}}\frac{\sqrt{h}}{\sqrt{\widetilde{h}}}
	=\sqrt{\frac{\varOmega^{2}-\varOmega\left\{\frac{\varGamma_{11}}{N^{2}}
			+\frac{2\varGamma_{12}\pounds_{\bm{u}}\psi}{N}
			+\varGamma_{22}\left[\left(\pounds_{\bm{u}}\psi\right)^{2}-\left(\mathrm{D}\psi\right)^{2}\right]\right\}
			-\frac{\varGamma_{11}\varGamma_{22}-\varGamma^{2}_{12}}{N^{2}}\left(\mathrm{D}\psi\right)^{2}}
		{\varOmega+\varGamma_{22}\left(\mathrm{D}\psi\right)^{2}}}.
\end{equation}
For any scalar, the first intrinsic covariant derivatives coincide,
$\widetilde{\mathrm D}_i\psi=\mathrm D_i\psi=\partial_i\psi$. 

The inverse spatial metric is
\begin{equation}
	\widetilde{h}^{ij}=\varOmega^{-1}\left[h^{ij}
	-\frac{\varGamma_{22}}{\varOmega+\varGamma_{22}\left(\mathrm{D}\psi\right)^{2}}
	\mathrm{D}^{i}\psi\mathrm{D}^{j}\psi\right].
\end{equation}
It follows that
\begin{equation}
	\widetilde{N}^{i}=\widetilde{h}^{ij}\widetilde{N}_{j}
	=N^{i}+W\mathrm{D}^{i}\psi,\label{eq:trans of Ni}
\end{equation}
where
\begin{equation}
	W\coloneqq\frac{\varGamma_{12}+\varGamma_{22}N\pounds_{\bm{u}}\psi}
	{\varOmega+\varGamma_{22}\left(\mathrm{D}\psi\right)^{2}}.
\end{equation}

The unit normals to the same constant-$\phi$ hypersurfaces are related by
\begin{equation}
	\widetilde{u}_{a}=-\widetilde N\nabla_a\phi=\varPhi u_a,
	\qquad
	\widetilde{u}^{a}=\varPhi^{-1}\left(u^{a}-\frac{W}{N}\mathrm D^{a}\psi\right),
\end{equation}
where $\mathrm D^a\psi\coloneqq h^{ab}\nabla_b\psi$. Hence the transformed extrinsic curvature is
\begin{equation}
	\widetilde{K}_{ij}=\frac{1}{2}\pounds_{\widetilde{\bm{u}}}\widetilde{h}_{ij}. \label{tldKij_def}
\end{equation}
Substituting Eqs.~(\ref{eq:trans of N})-(\ref{eq:trans of h_ij}) yields
\begin{align}
	\widetilde{K}_{ij}=\varPhi^{-1}\Biggl\{ & \varOmega K_{ij}
	-\frac{\varOmega W}{N}\mathrm{D}_{i}\mathrm{D}_{j}\psi
	-\frac{\varOmega}{N}\mathrm{D}_{(i}\psi\mathrm{D}_{j)}W
	+\frac{1}{2N}\left(N\pounds_{\bm{u}}\varOmega-W\mathrm{D}^{k}\psi\mathrm{D}_{k}\varOmega\right)h_{ij}\nonumber \\
	&+\frac{1}{2N}\left(N\pounds_{\bm{u}}\varGamma_{22}-W\mathrm{D}^{k}\psi\mathrm{D}_{k}\varGamma_{22}\right)
	\mathrm{D}_{i}\psi\mathrm{D}_{j}\psi
	+\frac{\varGamma_{22}}{N}\mathrm{D}_{(i}\psi\mathrm{D}_{j)}
	\left[N\pounds_{\bm{u}}\psi-W\left(\mathrm{D}\psi\right)^{2}\right]\Biggr\}. \label{tldKij_trans}
\end{align}
The definition of $W$ implies
\begin{equation}
	W\left[\varOmega+\varGamma_{22}\left(\mathrm{D}\psi\right)^{2}\right]
	=\varGamma_{12}+\varGamma_{22}N\pounds_{\bm{u}}\psi. \label{WGmmeql}
\end{equation}
Taking a spatial derivative gives
\begin{align}
	\varGamma_{22}\mathrm{D}_{i}\left(N\pounds_{\bm{u}}\psi\right)
	-\varGamma_{22}\mathrm{D}_{i}\left[W\left(\mathrm{D}\psi\right)^{2}\right]
	-\varOmega\mathrm{D}_{i}W
	=W\mathrm{D}_{i}\varOmega
	+W\left(\mathrm{D}\psi\right)^{2}\mathrm{D}_{i}\varGamma_{22}
	-N\pounds_{\bm{u}}\psi\mathrm{D}_{i}\varGamma_{22}
	-\mathrm{D}_{i}\varGamma_{12}.
\end{align}
Using this identity to eliminate $\mathrm D_iW$ in Eq.~(\ref{tldKij_trans}), we obtain
\begin{align}
	\widetilde{K}_{ij}=\varPhi^{-1}\Biggl\{ & \varOmega K_{ij}
	-\frac{\varOmega W}{N}\mathrm{D}_{i}\mathrm{D}_{j}\psi
	+\frac{1}{2N}\left(N\pounds_{\bm{u}}\varOmega-W\mathrm{D}^{k}\psi\mathrm{D}_{k}\varOmega\right)h_{ij}
	+\frac{1}{2N}\left(N\pounds_{\bm{u}}\varGamma_{22}-W\mathrm{D}^{k}\psi\mathrm{D}_{k}\varGamma_{22}\right)
	\mathrm{D}_{i}\psi\mathrm{D}_{j}\psi\nonumber \\
	&+\frac{1}{N}\mathrm{D}_{(i}\psi\Bigl[
	W\mathrm{D}_{j)}\varOmega
	+W\left(\mathrm{D}\psi\right)^{2}\mathrm{D}_{j)}\varGamma_{22}
	-N\pounds_{\bm{u}}\psi\mathrm{D}_{j)}\varGamma_{22}
	-\mathrm{D}_{j)}\varGamma_{12}\Bigr]\Biggr\}.\label{eq:trans of Kij}
\end{align}

The transformed acceleration is most naturally defined using the transformed intrinsic connection. Since the lapse function is a scalar, its first derivative is unchanged:
\begin{equation}
	\widetilde{a}_{i}\coloneqq\widetilde{\mathrm{D}}_{i}\ln\widetilde{N}
	=\mathrm{D}_{i}\ln\widetilde{N}
	=a_{i}+\varPhi^{-1}\mathrm{D}_{i}\varPhi.
\end{equation}

From the transformed normal vector,
\begin{equation}
	\pounds_{\widetilde{\bm{u}}}\psi
	=\varPhi^{-1}\left[\pounds_{\bm{u}}\psi
	-\frac{W}{N}\left(\mathrm{D}\psi\right)^{2}\right],
\end{equation}
and a second Lie derivative gives
\begin{align}
	\pounds^{2}_{\widetilde{\bm{u}}}\psi
	=\varPhi^{-2}\Biggl\{&\pounds^{2}_{\bm{u}}\psi
	-\frac{3W}{N}\mathrm{D}^{i}\psi\mathrm{D}_{i}(\pounds_{\bm{u}}\psi)
	+\frac{2W}{N}K^{ij}\mathrm{D}_{i}\psi\mathrm{D}_{j}\psi
	+\frac{2W^{2}}{N^{2}}\mathrm{D}^{i}\psi\mathrm{D}^{j}\psi\mathrm{D}_{i}\mathrm{D}_{j}\psi\nonumber\\
	&-\left(\mathrm D\psi\right)^2
	\left[\pounds_{\bm u}\left(\frac WN\right)
	-\frac WN\mathrm D^i\psi\mathrm D_i\left(\frac WN\right)\right]
	-\frac{2W}{N}\pounds_{\bm u}\psi\,a^i\mathrm D_i\psi\nonumber\\
	&-\left(\pounds_{\bm{u}}\ln\varPhi
	-\frac{W}{N}\mathrm{D}^{i}\psi\mathrm{D}_{i}\ln\varPhi\right)
	\left[\pounds_{\bm{u}}\psi
	-\frac{W}{N}(\mathrm{D}\psi)^{2}\right]\Biggr\}.\label{eq:trans of F}
\end{align}
Here the acceleration term follows from
$\pounds_{\bm u}(\mathrm D\psi)^2
=2\mathrm D^i\psi\mathrm D_i(\pounds_{\bm u}\psi)
-2K^{ij}\mathrm D_i\psi\mathrm D_j\psi
+2\pounds_{\bm u}\psi\,a^i\mathrm D_i\psi$.

The intrinsic Ricci tensor transforms according to
\begin{equation}
	\widetilde{R}_{ij}=R_{ij}+\mathrm{D}_{k}C^{k}_{ij}-\mathrm{D}_{j}C^{k}_{ik}
	+C^{k}_{kl}C^{l}_{ij}-C^{k}_{jl}C^{l}_{ik},
\end{equation}
where the difference between the two intrinsic Levi-Civita connections is
\begin{equation}
	C^{k}_{ij}\coloneqq\widetilde{\varGamma}^{k}_{ij}-\varGamma^{k}_{ij}
	=\frac{1}{2}\widetilde{h}^{kl}\left(
	\mathrm{D}_{i}\widetilde{h}_{jl}
	+\mathrm{D}_{j}\widetilde{h}_{il}
	-\mathrm{D}_{l}\widetilde{h}_{ij}\right). \label{Ckij_def}
\end{equation}
Note that the covariant derivative in (\ref{Ckij_def}) must be $\mathrm D_i$, i.e., the connection compatible with $h_{ij}$.
Substitution of (\ref{eq:trans of h_ij}) gives
\begin{align}
	C^{k}_{ij}= & \frac{1}{\varOmega}\delta^{k}_{(i}\mathrm{D}_{j)}\varOmega
	-\frac{1}{2\varOmega}h_{ij}\mathrm{D}^{k}\varOmega
	+\frac{\varGamma_{22}}{\varOmega+\varGamma_{22}(\mathrm{D}\psi)^{2}}
	\mathrm{D}^{k}\psi\mathrm{D}_{i}\mathrm{D}_{j}\psi
	+\frac{1}{\varOmega+\varGamma_{22}(\mathrm{D}\psi)^{2}}
	\mathrm{D}^{k}\psi\mathrm{D}_{(i}\psi\mathrm{D}_{j)}\varGamma_{22}\nonumber \\
	&-\frac{\varGamma_{22}}{\varOmega\left[\varOmega+\varGamma_{22}(\mathrm{D}\psi)^{2}\right]}
	\mathrm{D}^{k}\psi\mathrm{D}_{(i}\psi\mathrm{D}_{j)}\varOmega
	+\frac{\varGamma_{22}}{2\varOmega\left[\varOmega+\varGamma_{22}(\mathrm{D}\psi)^{2}\right]}
	h_{ij}\mathrm{D}^{k}\psi\left(\mathrm{D}^{l}\psi\mathrm{D}_{l}\varOmega\right)\nonumber \\
	&-\frac{1}{2\varOmega}\mathrm{D}_{i}\psi\mathrm{D}_{j}\psi\mathrm{D}^{k}\varGamma_{22}
	+\frac{\varGamma_{22}}{2\varOmega\left[\varOmega+\varGamma_{22}(\mathrm{D}\psi)^{2}\right]}
	\mathrm{D}_{i}\psi\mathrm{D}_{j}\psi\mathrm{D}^{k}\psi
	\left(\mathrm{D}^{l}\psi\mathrm{D}_{l}\varGamma_{22}\right).
\end{align}

Equations~(\ref{eq:trans of N})-(\ref{eq:trans of h_ij}) may alternatively be taken as the definition of a broader spatially covariant field transformation,
\begin{align}
	\widetilde{N}&=\varPhi N,\label{eq:trans of N-1}
	\\
	\widetilde{N}_{i}&=\varPsi N_{i}+\varLambda\mathrm{D}_{i}\psi,\label{eq:trans of N_i-1}
	\\
	\widetilde{h}_{ij}&=\varOmega h_{ij}+\varGamma\mathrm{D}_{i}\psi\mathrm{D}_{j}\psi,\label{eq:trans of h_ij-1}
\end{align}
where $\varPhi$, $\varPsi$, $\varLambda$, $\varOmega$, and $\varGamma$ are spatial scalars constructed from
\begin{equation}
	t,N,N_{i},h_{ij},K_{ij},{}^{3}\!R_{ij},\psi,
	\pounds_{\bm{u}}\psi,\pounds^{2}_{\bm{u}}\psi
\end{equation}
and their spatial derivatives, with all indices contracted appropriately. When the coefficients have the dependence inherited from Eq.~(\ref{eq:disformal trans}), the transformation is algebraic in the metric and involves at most first derivatives of the scalar fields. Dependence on $K_{ij}$, $\pounds^{2}_{\bm u}\psi$, or their spatial derivatives makes the map derivative-dependent, so the Jacobian criterion derived above no longer establishes either the invertibility of the map or the preservation of the number of degrees of freedom. We therefore restrict the following discussion to the covariant transformation in Eq.~(\ref{eq:disformal trans}).

\subsection{A restricted disformal transformation}

We now examine whether an action in the class (\ref{eq:S0}) is mapped into the same class. A generic transformation fails this test. In particular, the terms $\pounds_{\bm u}\varOmega$ and $\pounds_{\bm u}\varGamma_{22}$ in Eq.~(\ref{eq:trans of Kij}), as well as $\pounds_{\bm u}\ln\varPhi$ and $\pounds_{\bm u}(W/N)$ in Eq.~(\ref{eq:trans of F}), generate $\pounds_{\bm u}N$ whenever the transformation functions have generic lapse dependence. Since Eq.~(\ref{eq:S0}) excludes the velocity of the lapse, the class of Lagrangians defined there is not closed under the general transformation: the image of an action in the class need not be expressible in the form (\ref{eq:S0}).

One simple restricted subclass, sufficient for closure but not claimed to exhaust all possible cancellations, is defined by
\begin{itemize}
	\item $\varGamma_{11}=\varGamma_{12}=0$;
	\item $\varOmega=\varOmega(\phi,\psi,X^{22})$ and
	$\varGamma_{22}\eqqcolon\varGamma=\varGamma(\phi,\psi,X^{22})$, with no dependence on $X^{11}$ or $X^{12}$.
\end{itemize}
In the unitary gauge these functions depend on $t$, $\psi$, and $Y=X^{22}$, but have no independent lapse dependence. The derivative dependence of this subclass is therefore entirely through $\nabla_a\psi$. An explicit dependence on $\phi=t$ is still allowed. Consequently, the transformed building blocks contain no $\pounds_{\bm u}N$.

For this subclass, define 
\begin{equation}
	w \coloneqq\frac{W}{N}
	=\frac{\varGamma\pounds_{\bm{u}}\psi}
	{\varOmega+\varGamma(\mathrm{D}\psi)^{2}},
	\qquad
	\varPhi=\sqrt{\varOmega\left[1-
		\frac{\varGamma\left(\pounds_{\bm{u}}\psi\right)^{2}}
		{\varOmega+\varGamma\left(\mathrm{D}\psi\right)^{2}}\right]}.
\end{equation}
The basic transformed quantities reduce to
\begin{align}
	\widetilde{K}_{ij}&=\varPhi^{-1}\Biggl[  \varOmega K_{ij}
	-w \varOmega\mathrm{D}_{i}\mathrm{D}_{j}\psi
	+\frac{1}{2}\left(\pounds_{\bm{u}}\varOmega
	-w\mathrm{D}^{k}\psi\mathrm{D}_{k}\varOmega\right)h_{ij}\nonumber \\
	&\quad 
    +\frac{1}{2}\left(\pounds_{\bm{u}}\varGamma
	-w\mathrm{D}^{k}\psi\mathrm{D}_{k}\varGamma\right)
	\mathrm{D}_{i}\psi\mathrm{D}_{j}\psi
	+\frac{\pounds_{\bm{u}}\psi}{\varOmega+\varGamma(\mathrm{D}\psi)^{2}}
	\mathrm{D}_{(i}\psi\left(\varGamma\mathrm{D}_{j)}\varOmega
	-\varOmega\mathrm{D}_{j)}\varGamma\right)\Biggr],
	\\
	\pounds_{\widetilde{\bm{u}}}\psi
	&=\varPhi^{-1}\left[\pounds_{\bm{u}}\psi
	-w\left(\mathrm{D}\psi\right)^{2}\right],
	\\
	\pounds^{2}_{\widetilde{\bm{u}}}\psi
	&=\varPhi^{-2}\Biggl\{\pounds^{2}_{\bm{u}}\psi
	-3w\mathrm{D}^{i}\psi\mathrm{D}_{i}(\pounds_{\bm{u}}\psi)
	+2w K^{ij}\mathrm{D}_{i}\psi\mathrm{D}_{j}\psi
	+2w^{2}\mathrm{D}^{i}\psi\mathrm{D}^{j}\psi\mathrm{D}_{i}\mathrm{D}_{j}\psi\nonumber \\
	&\quad 
    -\left(\mathrm{D}\psi\right)^{2}
	\left(\pounds_{\bm{u}}w
	-w\mathrm{D}^{i}\psi\mathrm{D}_{i}w\right)
	-2w\pounds_{\bm u}\psi\,a^i\mathrm D_i\psi\nonumber\\
	&\quad
    -\left(\pounds_{\bm{u}}\ln\varPhi
	-w\mathrm{D}^{i}\psi\mathrm{D}_{i}\ln\varPhi\right)
	\left[\pounds_{\bm{u}}\psi
	-w(\mathrm{D}\psi)^{2}\right]\Biggr\}.
\end{align}
All quantities on the right-hand sides belong to the set of building blocks admitted in (\ref{eq:S0}), including their spatial derivatives. Hence the class (\ref{eq:S0}) is closed under this restricted transformation, provided the metric and Jacobian regularity conditions derived above hold.

\section{Conclusion} \label{sec:con}

We have proposed a novel approach to constructing higher-derivative bi-scalar-tensor theories within the framework of spatially covariant gravity. We have identified the conditions required to eliminate the unwanted scalar mode, so that the theory propagates four degrees of freedom. 

The starting point is the spatially covariant scalar field action (\ref{eq:S0}). One scalar field $\phi$ defines the foliation and is put in unitary gauge, while a second field $\psi$ remains explicit and may enter through $\pounds_{\bm{u}}^{2}\psi$ and arbitrary spatial derivatives. Upon restoration of temporal diffeomorphism invariance, this action can be interpreted as a generally covariant bi-scalar-tensor theory. This formulation allows higher spatial derivatives while keeping explicit control over time derivatives, but it does not make a general action of the form (\ref{eq:S0}) ghost-free. Health follows only after the relevant degeneracy and consistency conditions, together with the assumptions used to derive them, have been imposed.

To perform the nonlinear analysis without inverting a general nonlinear relation between velocities and momenta, we introduced the auxiliary variables $A$, $F$, and $B_{ij}$ and rewrote the action in the first-order form (\ref{model}). The primary and secondary constraints and their algebra, summarized by the operator $\mathcal P^{\alpha\beta}$ in Eq.~(\ref{eq:matrix}) and developed in Sec.~\ref{subsec:Constraint-algebra}, show that a generic nondegenerate theory has six first-class and sixteen second-class constraints in a 38-dimensional phase space. It therefore propagates five physical degrees of freedom. Besides the two tensor polarizations, the scalar sector contains the mode encoded in the SCG metric variables, the expected mode of $\psi$, and an additional mode originating from the higher-time-derivative dependence on $\psi$. In the nondegenerate branch this last mode is the would-be Ostrogradsky degree of freedom that the subsequent constraints are designed to remove.

To derive the degeneracy condition, we restricted attention to the branch in which the lapse Hessian $\delta^{2}S/\delta N\delta N$ in \eqref{PB:calCpiN} and the effective tensor block $\mathcal J^{ij,kl}$ in \eqref{calJ} are invertible, and $\mathcal P^{\alpha\beta}$ has exactly one zero mode. Eliminating the lapse and tensor components of the corresponding zero mode gives the degeneracy condition (\ref{eq:degeneracy condition}) with $\mathcal{D}(\vec{x},\vec{y})$ given in \eqref{eq:degeneracy condition function}. This condition produces the tertiary constraint $\theta$ through Eq.~(\ref{eq:new constraint}), but degeneracy alone may leave a residual half degree of freedom. To fully eliminate the unwanted mode, an additional condition is necessary. Projecting the remaining primary-tertiary bracket yields Eq.~(\ref{eq:consistency condition_0}) and motivates the additional consistency condition (\ref{eq:consistency condition}) with $\mathcal{F}(\vec{x},\vec{y})$ given in \eqref{eq:conscalFxpl}. 

After both conditions are imposed, the constraint chain can close in two ways. In Case I, the preservation of $\theta$ is automatic. The system has seven first-class and sixteen second-class constraints, as summarized in Table~\ref{tab:PBcase1}. In Case II, the same preservation equation generates the quaternary constraint $\omega$ in Eq.~(\ref{eq:quaternary constraint}). Provided the residual operator that fixes the remaining multiplier is invertible, there are six first-class and eighteen second-class constraints, as summarized in Table~\ref{tab:PBcase2}. Both cases propagate four physical degrees of freedom, corresponding to two tensor and two scalar modes. Thus the four-mode result is not a property of the general spatially covariant scalar field action, but of the subclasses satisfying Eqs.~(\ref{eq:degeneracy condition}) and (\ref{eq:consistency condition}) under the assumed invertibility, rank, regularity, and boundary conditions.

The degeneracy and consistency conditions derived in this work characterize only the branch specified by the above invertibility and corank-one assumptions. Other branches can arise if the lapse Hessian or $\mathcal J^{ij,kl}$ is singular, or if the relevant operator has more than one zero mode. The bi-Galileon example, whose unitary-gauge form is displayed in Appendix~\ref{app:bigalug}, illustrates this point. It belongs to the broad class defined by Eq.~(\ref{eq:S0}), but $F$ is linear and has no kinetic mixing with $B_{ij}$, so the elimination scheme used in the main analysis need not apply. It should instead be treated by a constraint analysis adapted to a different invertible block. Consequently, the fact that a theory belongs to the broad class or is known to be healthy does not imply that it must satisfy the two conditions derived in our chosen branch. Distinct degeneracy mechanisms can lead to the same physical number of degrees of freedom \citep{BouzariNezhad:2026zsj}.

We have also examined the two-field disformal transformation (\ref{eq:disformal trans}). We obtained the inverse metric and volume element, including the field-space determinant in Eq.~(\ref{eq:trans fo =00005Csqrtg}), derived the necessary and sufficient Jacobian criterion for local invertibility in the bi-scalar case, and expressed the transformed ADM variables in Eqs.~(\ref{eq:trans of N})-(\ref{eq:trans of h_ij}) together with the transformed $K_{ij}$ and $\pounds_{\bm{u}}^{2}\psi$ in Eqs.~(\ref{eq:trans of Kij}) and (\ref{eq:trans of F}). A generic covariant two-field transformation generates $\pounds_{\bm{u}}N$ and therefore takes an action of the form (\ref{eq:S0}) outside the framework analyzed here. For the transformations evaluated in this work, closure is retained by setting $\varGamma_{11}=\varGamma_{12}=0$ and taking $\varOmega$ and $\varGamma_{22}$ to be independent of $N$ (equivalently, independent of $X^{11}$ and $X^{12}$ in unitary gauge). An invertible and regular transformation in this restricted class preserves the number of physical degrees of freedom, but it does not follow that the transformed action realizes the particular degeneracy and consistency branch of Eqs.~(\ref{eq:degeneracy condition}) and (\ref{eq:consistency condition}) \citep{Watanabe:2015uqa,Domenech:2025gao}.

Several questions remain open. It will be interesting to classify the branches excluded by the invertibility assumptions in this work. It is also important to apply the functional conditions (\ref{eq:degeneracy condition}) and (\ref{eq:consistency condition}) to derive explicit families of Lagrangians, and to analyze their features in cosmological perturbations and gravitational waves. The relation to the symmetric interacting-hypersurface construction reviewed in Appendix~\ref{app:dec} also deserves a full nonlinear constraint analysis \citep{Yu:2024sed}. Finally, allowing a controlled velocity of the lapse, enlarging the construction to more explicit scalar fields, and determining which multi-field disformal transformations preserve each constraint branch may broaden the theory space \citep{Gao:2018znj,Gao:2019lpz,BouzariNezhad:2026zsj}. We will explore these issues in future work.

\begin{acknowledgments}
	X. G. was supported by the National Natural Science Foundation of China (NSFC) under Grant Nos. 12475068 and 11975020 and by the Guangdong Basic and Applied Basic Research Foundation under Grant No. 2025A1515012977. The work of T. K. was supported by JSPS KAKENHI Grant No.~JP25K07308.
    We used Claude Fable 5 and ChatGPT-5.6 to cross-check our Hamiltonian analysis.
\end{acknowledgments}

\appendix

\section{Preservation of $\mathcal{C}_{i}$} \label{app:cc_calCi}

Since the Poisson brackets of $\mathcal{C}_{i}$ with all primary
constraints vanish weakly and $\mathcal{C}_{i}$ has no explicit time
dependence, its preservation equation reduces to
\begin{equation}
	\dot{\mathcal C}_{i}(\vec{x})
	\approx
	\mathfrak{D}_{0}\mathcal C_{i}(\vec{x})
	+\sum_{A}\int\mathrm{d}^{3}y\,
	\lambda_{A}(\vec{y})
	[\mathcal C_{i}(\vec{x}),\varphi^{A}(\vec{y})]
	\approx
	\mathfrak{D}_{0}\mathcal C_{i}(\vec{x})
	=
	[\mathcal C_{i}(\vec{x}),H_{0}]
	\approx0.
\end{equation}
We must check whether this equation leads to a tertiary constraint. We find
\begin{align}
	\left[\mathcal{C}_{i}\left(\vec{x}\right),H_{0}\right] & =\int\mathrm{d}^{3}y\left\{ \left[\mathcal{C}_{i}\left(\vec{x}\right),N\left(\vec{y}\right)C\left(\vec{y}\right)+N^{j}\left(\vec{y}\right)\mathcal{C}_{j}\left(\vec{y}\right)\right]\right\} \nonumber \\
	& \approx\int\mathrm{d}^{3}y\left\{ N\left(\vec{y}\right)\left[\mathcal{C}_{i}\left(\vec{x}\right),C\left(\vec{y}\right)\right]+\left[\mathcal{C}_{i}\left(\vec{x}\right),N\left(\vec{y}\right)\right]C\left(\vec{y}\right)\right\} ,\label{eq:=00003D00005Bci.H0=00003D00005D}
\end{align}
where in the last step we used Eq.~\eqref{PB_calCi_Q}.
Using Eq.~\eqref{PB_calCi_Q_proven}, we obtain
\begin{align}
	\int\mathrm{d}^{3}x\mathrm{d}^{3}y\left.\xi^{i}\left(\vec{x}\right)f\left(\vec{y}\right)\left[\mathcal{C}_{i}\left(\vec{x}\right),N\left(\vec{y}\right)\right]\right. & \approx\int\mathrm{d}^{3}x\left.N\left(\vec{x}\right)\pounds_{\vec{\xi}}f\left(\vec{x}\right)\right.\nonumber \\
	& =-\int\mathrm{d}^{3}x\left.f\left(\vec{x}\right)\pounds_{\vec{\xi}}N\left(\vec{x}\right)\right.\nonumber \\
	& =-\int\mathrm{d}^{3}x\mathrm{d}^{3}y\left.f\left(\vec{y}\right)\delta^{(3)}\left(\vec{y}-\vec{x}\right)\xi^{i}\left(\vec{x}\right)\partial_{x^{i}}N\left(\vec{x}\right)\right..
\end{align}
In the second equality, we used
$\pounds_{\vec{\xi}}f=\partial_i(f\xi^i)$ because the test function $f$
is a scalar density of weight one, so that
$\int\mathrm d^3x\,Nf$ is invariant under time-independent spatial
diffeomorphisms (equivalently, $f/\sqrt h$ is a scalar). Therefore,
\begin{align}
	\left[\mathcal{C}_{i}\left(\vec{x}\right),N\left(\vec{y}\right)\right] & \approx-\delta^{(3)}\left(\vec{y}-\vec{x}\right)\partial_{x^{i}}N\left(\vec{x}\right).
\end{align}

Similarly,
\begin{align}
	\int\mathrm{d}^{3}x\mathrm{d}^{3}y\left.\xi^{i}\left(\vec{x}\right)f\left(\vec{y}\right)\left[\mathcal{C}_{i}\left(\vec{x}\right),C\left(\vec{y}\right)\right]\right. & \approx\int\mathrm{d}^{3}y\left.C\left(\vec{y}\right)\pounds_{\vec{\xi}}f\left(\vec{y}\right)\right.\nonumber \\
	& =\int\mathrm{d}^{3}y\left.C\left(\vec{y}\right)\xi^{i}\left(\vec{y}\right)\partial_{y^{i}}f\left(\vec{y}\right)\right.\nonumber \\
	& =\int\mathrm{d}^{3}x\mathrm{d}^{3}y\left.\xi^{i}\left(\vec{x}\right)\delta^{(3)}\left(\vec{x}-\vec{y}\right)C\left(\vec{y}\right)\partial_{y^{i}}f\left(\vec{y}\right)\right..
\end{align}
Since $C$ is a spatial density of unit weight, $f$ is now taken to be a
spatial scalar. Thus,
\begin{align}
	\int\mathrm{d}^{3}y\left.f\left(\vec{y}\right)\left[\mathcal{C}_{i}\left(\vec{x}\right),C\left(\vec{y}\right)\right]\right. & \approx\int\mathrm{d}^{3}y\left.\delta^{(3)}\left(\vec{x}-\vec{y}\right)C\left(\vec{y}\right)\partial_{y^{i}}f\left(\vec{y}\right)\right.\nonumber \\
	& \simeq-\int\mathrm{d}^{3}y\left.f\left(\vec{y}\right)\partial_{y^{i}}\left[\delta^{(3)}\left(\vec{x}-\vec{y}\right)C\left(\vec{y}\right)\right]\right..
\end{align}
Comparing the two sides gives
\begin{equation}
	\left[\mathcal{C}_{i}\left(\vec{x}\right),C\left(\vec{y}\right)\right]\approx-\partial_{y^{i}}\left[\delta^{(3)}\left(\vec{x}-\vec{y}\right)C\left(\vec{y}\right)\right].
\end{equation}

Finally, substituting these brackets into
Eq.~\eqref{eq:=00003D00005Bci.H0=00003D00005D}, we obtain
\begin{align}
	\left[\mathcal{C}_{i}\left(\vec{x}\right),H_{0}\right] & \approx\int\mathrm{d}^{3}y\left\{ -\partial_{y^{i}}\left[\delta^{(3)}\left(\vec{x}-\vec{y}\right)C\left(\vec{y}\right)\right]N\left(\vec{y}\right)-\delta^{(3)}\left(\vec{y}-\vec{x}\right)\partial_{x^{i}}N\left(\vec{x}\right)C\left(\vec{y}\right)\right\} \nonumber \\
	& \simeq\int\mathrm{d}^{3}y\left.\delta^{(3)}\left(\vec{y}-\vec{x}\right)\left\{ C\left(\vec{y}\right)\partial_{y^{i}}N\left(\vec{y}\right)-\partial_{x^{i}}N\left(\vec{x}\right)C\left(\vec{y}\right)\right\} \right.=0.
\end{align}
As expected, the preservation equation for $\mathcal{C}_{i}$ is automatically satisfied and therefore generates no tertiary constraint.

\section{Bi-Galileon in the unitary gauge} \label{app:bigalug}

To illustrate the scope of the broad action (\ref{eq:S0}), we consider the
bi-Galileon theory with two scalar fields, $\phi$ and $\psi$. We show that, in
the unitary gauge $\phi=t$ and up to boundary terms, its Lagrangian belongs to
the class defined by Eq.~\eqref{eq:S0}.

The bi-Galileon Lagrangian containing terms at most quadratic in second
derivatives of the scalar fields takes the form \cite{Padilla:2012dx,Kobayashi:2013ina}
\begin{equation}
	\mathcal{L}=G_{2}(X^{IJ},\phi^{K})-G_{3L}(X^{IJ},\phi^{K})\square\phi^{L}+G_{4}(X^{IJ},\phi^{K})R+G_{4,\langle IJ\rangle}\left(\square\phi^{I}\square\phi^{J}-\nabla_{\mu}\nabla_{\nu}\phi^{I}\nabla^{\mu}\nabla^{\nu}\phi^{J}\right),\label{eq:Bi-Galileon}
\end{equation}
where $\phi^{I}=\left\{ \phi,\psi\right\}$. Here
$X^{IJ}\coloneqq-\frac{1}{2}g^{\mu\nu}\nabla_{\mu}\phi^{I}\nabla_{\nu}\phi^{J}$
and
$G_{4,\langle IJ\rangle}\coloneqq\frac{1}{2}\left(\partial G_{4}/\partial X^{IJ}+\partial G_{4}/\partial X^{JI}\right)$.
Provided that the standard total-symmetry conditions on
$G_{3I,\langle JK\rangle}$ and
$G_{4,\langle IJ\rangle,\langle KL\rangle}$ hold, the field equations are of
second order. A generic nondegenerate theory in this class propagates two
tensor and two scalar degrees of freedom.

In the unitary gauge $\phi=t$, the $3+1$ decomposition of
Eq.~\eqref{eq:Bi-Galileon} is
\begin{align}
	\mathcal{L}_{3+1} & =G_{2}-G_{31}\left(\frac{\pounds_{\bm{u}}N}{N^{2}}-\frac{K}{N}\right)-G_{32}(\mathrm{D}^{2}\psi-K\pounds_{\bm{u}}\psi-\pounds^{2}_{\bm{u}}\psi+a^{i}\mathrm{D}_{i}\psi)\nonumber \\
	& \quad+\mathcal{A}_{K}\left(K^{2}-K_{ij}K^{ij}\right)+G_{4}R^{(3)}-2a^{i}\mathrm{D}_{i}G_{4}-2G_{4,\phi}\frac{K}{N}-2G_{4,\psi}K\pounds_{\bm{u}}\psi\nonumber \\
	& \quad+G_{4,\langle 11 \rangle}\left[\frac{2a_{i}a^{i}}{N^{2}}\right]\nonumber \\
	& \quad+2G_{4,\langle 12 \rangle}\left[\frac{1}{N}(K_{ij}\mathrm{D}^{i}\mathrm{D}^{j}\psi-K\mathrm{D}^{2}\psi)+\frac{\pounds_{\bm{u}}N}{N^{2}}\mathrm{D}^{2}\psi-\frac{K}{N}a^{i}\mathrm{D}_{i}\psi-\frac{2}{N}a^{i}\mathrm{D}_{i}(\pounds_{\bm{u}}\psi)+\frac{2}{N}a_{i}K^{ij}\mathrm{D}_{j}\psi\right]\nonumber \\
	& \quad+G_{4,\langle 22 \rangle}\left[((\mathrm{D}^{2}\psi)^{2}-(\mathrm{D}_{i}\mathrm{D}_{j}\psi)^{2})+2(\pounds_{\bm{u}}\psi)(K_{ij}\mathrm{D}^{i}\mathrm{D}^{j}\psi-K\mathrm{D}^{2}\psi)-2\mathrm{D}^{2}\psi\pounds^{2}_{\bm{u}}\psi -2KK^{ij}\mathrm{D}_i \psi\mathrm{D}_j \psi\right.\nonumber \\
	& \quad\quad\quad\left.+2a^{i}\mathrm{D}_{i}\psi\mathrm{D}^{2}\psi+2\mathrm{D}_{i}(\pounds_{\bm{u}}\psi)\mathrm{D}^{i}(\pounds_{\bm{u}}\psi)-4K^{ij}\mathrm{D}_{i}(\pounds_{\bm{u}}\psi)\mathrm{D}_{j}\psi+2K_{ik}K^{il}\mathrm{D}^{k}\psi\mathrm{D}_{l}\psi+2K\mathrm{D}^{i}\psi\mathrm{D}_{i}(\pounds_{\bm{u}}\psi)\right],\label{eq:bi-Galileon ADM}
\end{align}
with 
\begin{equation}
	\mathcal{A}_{K}=G_{4}+G_{4,\langle 11 \rangle }\frac{1}{N^{2}}+G_{4,\langle 12 \rangle}\frac{2}{N}\pounds_{\bm{u}}\psi+G_{4,\langle 22 \rangle}(\pounds_{\bm{u}}\psi)^{2}.
\end{equation}

The terms involving $\pounds_{\bm{u}}N$ can be eliminated by adding a
boundary term and integrating by parts:
\begin{align}
	S & =\int\mathrm{d}t\mathrm{d}^{3}x\,N\sqrt{h}\mathcal{L} \simeq\int\mathrm{d}t\mathrm{d}^{3}x\,\left(N\sqrt{h}\mathcal{L}+\partial_{t}\left(\sqrt{h}F\right)-\sqrt{h}\mathrm{D}_{i}\left(N^{i}F\right)\right)\nonumber \\
	& =\int\mathrm{d}t\mathrm{d}^{3}x\,N\sqrt{h}\left(\mathcal{L}+FK+\pounds_{\bm{u}}F\right),
\end{align}
where $\simeq$ denotes equality up to a boundary term, and $F$ is a boundary-term function unrelated to the auxiliary field
denoted by the same symbol in Sec.~\ref{sec:scgsf}. 
We choose
$F\coloneqq F_{3}+F_{4}\mathrm{D}^{2}\psi$, where $F_{3}$ and $F_{4}$ are
functions of $t,N,\psi,\pounds_{\bm{u}}\psi,Z$, with
$Z\coloneqq \mathrm{D}_{i}\psi\,\mathrm{D}^{i}\psi$. Then
\begin{align}
	\pounds_{\bm{u}}F_{3} & =\frac{\partial F_{3}}{\partial t}\frac{1}{N}+\frac{\partial F_{3}}{\partial N}\pounds_{\bm{u}}N+\frac{\partial F_{3}}{\partial\psi}\pounds_{\bm{u}}\psi+\frac{\partial F_{3}}{\partial\left(\pounds_{\bm{u}}\psi\right)}\pounds^{2}_{\bm{u}}\psi+\frac{\partial F_{3}}{\partial Z}\pounds_{\bm{u}}Z,\\
	\pounds_{\bm{u}}F_{4} & =\frac{\partial F_{4}}{\partial t}\frac{1}{N}+\frac{\partial F_{4}}{\partial N}\pounds_{\bm{u}}N+\frac{\partial F_{4}}{\partial\psi}\pounds_{\bm{u}}\psi+\frac{\partial F_{4}}{\partial\left(\pounds_{\bm{u}}\psi\right)}\pounds^{2}_{\bm{u}}\psi+\frac{\partial F_{4}}{\partial Z}\pounds_{\bm{u}}Z,
\end{align}
where
\begin{equation}
	\pounds_{\bm{u}}Z
	=
	2\mathrm{D}^{i}\psi\,
	\mathrm{D}_{i}
	\left(\pounds_{\bm{u}}\psi\right)
	-2K^{ij}
	\mathrm{D}_{i}\psi\,\mathrm{D}_{j}\psi
	+2\frac{\mathrm{D}_i N}{N}\mathrm{D}^{i}\psi\pounds_{\bm{u}}\psi
	.
\end{equation}
Choosing $F_{3}$ and $F_{4}$ to satisfy
\begin{equation}
	\frac{\partial F_{3}}{\partial N}=\frac{G_{31}}{N^{2}},\quad\frac{\partial F_{4}}{\partial N}=-\frac{2G_{4,\langle 12 \rangle}}{N^{2}},
\end{equation}
eliminates the terms proportional to $\pounds_{\bm{u}}N$. Expanding
$\pounds_{\bm{u}}F$ introduces terms with three derivatives in total, for
example,
\begin{align}
	F_{4}\pounds_{\bm{u}}(\mathrm{D}^{2}\psi)= & F_{4}\Biggl[\mathrm{D}^{2}(\pounds_{\bm{u}}\psi)+2a^{i}\mathrm{D}_{i}(\pounds_{\bm{u}}\psi)+\left(\mathrm{D}^{i}a_{i}+a^{i}a_{i}\right)\pounds_{\bm{u}}\psi\nonumber \\
	& -2K^{ij}\mathrm{D}_{i}\mathrm{D}_{j}\psi-\left(2\mathrm{D}_{i}K^{ij}-\mathrm{D}^{j}K+2a_{i}K^{ij}-Ka^{j}\right)\mathrm{D}_{j}\psi\Biggr].
\end{align}
These terms contain only spatial derivatives of the building blocks admitted
in Eq.~\eqref{eq:S0} and hence remain within the broad class considered here.

After these integrations by parts, Eq.~\eqref{eq:bi-Galileon ADM} can be
organized as
\begin{equation}
	\mathcal{L}_{3+1}=\mathcal{L}_{0}+\mathcal{L}_{1}+\mathcal{L}_{2}+\mathcal{L}_{3},\label{eq:Bi-Galileon in unitary}
\end{equation}
with 
\begin{align}
	\mathcal{L}_{0}&=G_{2}+\frac{1}{N}F_{3,t}+F_{3,\psi}\pounds_{\bm{u}}\psi,
	\\
	\mathcal{L}_{1}&=  b_{1}K+b_{2}K^{ij}\mathrm{D}_{i}\psi\mathrm{D}_{j}\psi+b_{3}a^{i}\mathrm{D}_{i}\psi+b_{4}\mathrm{D}^{2}\psi+b_{5}\mathrm{D}^{i}\psi\mathrm{D}_{i}(\pounds_{\bm{u}}\psi)+b_{6}\pounds^{2}_{\bm{u}}\psi,
	\label{eq:L1}
	\\ 
	\mathcal{L}_{2}&=  c_{1}K_{ij}K^{ij}+c_{2}K_{ij}K^{jk}\mathrm{D}^{i}\psi\mathrm{D}_{k}\psi+c_{3}a_{i}a^{i}+c_{4}\mathrm{D}_{i}\mathrm{D}_{j}\psi\mathrm{D}^{i}\mathrm{D}^{j}\psi+c_{5}\mathrm{D}_{i}(\pounds_{\bm{u}}\psi)\mathrm{D}^{i}(\pounds_{\bm{u}}\psi)\nonumber \\
	&\quad  +c_{6}K_{ij}a^{i}\mathrm{D}^{j}\psi+c_{7}K_{ij}\mathrm{D}^{i}\mathrm{D}^{j}\psi+c_{8}K_{ij}\mathrm{D}^{j}\mathrm{D}^{k}\psi\mathrm{D}^{i}\psi\mathrm{D}_{k}\psi+c_{9}K_{ij}\mathrm{D}^{i}(\pounds_{\bm{u}}\psi)\mathrm{D}^{j}\psi\nonumber \\
	&\quad 
	+c_{10}a^{i}\mathrm{D}_{i}\mathrm{D}_{j}\psi\mathrm{D}^{j}\psi+c_{11}a^{i}\mathrm{D}_{i}(\pounds_{\bm{u}}\psi)+c_{12}\mathrm{D}_{i}\mathrm{D}_{j}\psi\mathrm{D}^{i}(\pounds_{\bm{u}}\psi)\mathrm{D}^{j}\psi\nonumber\\
	&\quad +c_{13}K K^{ij}\mathrm{D}_{i}\psi\mathrm{D}_{j}\psi+c_{14}K^{ij}\mathrm{D}_{i}\psi\,\mathrm{D}_{j}\psi\mathrm{D}^{2}\psi+c_{15}R^{(3)},\label{eq:L2}
	\\
	\mathcal{L}_{3}&=  d_{1}K^{2}+d_{2}K\mathrm{D}^{2}\psi+d_{3}K\mathrm{D}^{i}\mathrm{D}^{j}\psi\mathrm{D}_{i}\psi\mathrm{D}_{j}\psi+d_{4}K\mathrm{D}^{i}\psi\mathrm{D}_{i}(\pounds_{\bm{u}}\psi)\nonumber \\
	&\quad 
	+d_{5}a^{i}\mathrm{D}_{i}\psi\mathrm{D}^{2}\psi+d_{6}(\mathrm{D}^{2}\psi)^{2}+d_{7}\mathrm{D}^{2}\psi\mathrm{D}^{i}\psi\mathrm{D}_{i}(\pounds_{\bm{u}}\psi)+d_{8}\mathrm{D}^{2}\psi\pounds^{2}_{\bm{u}}\psi.\label{eq:L3}
\end{align}
The coefficients $b_i$, $c_i$, and $d_i$ in Eqs.~\eqref{eq:L1}-\eqref{eq:L3}
obey
\begin{align}
	d_{1}=-c_{1}
	&=
	G_{4}
	+\frac{G_{4,\langle 11\rangle}}{N^{2}}
	+\frac{2G_{4,\langle 12\rangle}}{N}
	\pounds_{\bm{u}}\psi
	+G_{4,\langle 22\rangle}
	\left(\pounds_{\bm{u}}\psi\right)^{2},
	\\
	c_{2}=-2c_{4}=2d_{6}
	&=
	2G_{4,\langle 22\rangle},
	\\
	c_{7}=-d_{2}
	&=
	\frac{2G_{4,\langle 12\rangle}}{N}
	+2G_{4,\langle 22\rangle}
	\pounds_{\bm{u}}\psi,
	\\
	c_{8}=-2c_{12}=-2d_{3}=2d_{7}
	&=
	4F_{4,Z},
	\\
	c_{5}=d_{4}=-d_{8}
	&=
	2G_{4,\langle 22\rangle}
	-F_{4,\pounds_{\bm{u}}\psi}.
\end{align}
The remaining coefficients are
\begin{align}
    b_{1}
	&=
	\frac{G_{31}-2G_{4,\phi}}{N}
	+\left(
	G_{32}-2G_{4,\psi}
	\right)
	\pounds_{\bm{u}}\psi
	+F_{3}
	-ZF_{4,\psi},
    \qquad 
    b_{2}
	=
	-2F_{3,Z}
	+2F_{4,\psi},
    \notag \\
    b_{3}
	&=
	-G_{32}
	-2G_{4,\psi}
	+\left(
	2F_{3,Z}-F_{4,\psi}
	\right)
	\pounds_{\bm{u}}\psi,
    \qquad 
    b_{4}
	=
	-G_{32}
	+\frac{F_{4,t}}{N}
	+F_{4,\psi}\pounds_{\bm{u}}\psi,
    \qquad 
    b_{5}
	=
	2F_{3,Z}
	-F_{4,\psi},
    \notag \\  
    b_{6}
	&=
	G_{32}
	+F_{3,\pounds_{\bm{u}}\psi},
    \qquad 
    c_{3}
	=
	\frac{4G_{4,\langle 11\rangle}}{N^{2}}
	+\frac{4G_{4,\langle 12\rangle}}{N}
	\pounds_{\bm{u}}\psi,
    \qquad 
    c_6=0,
    \qquad 
    c_{9}
	=
	-4G_{4,\langle 22\rangle}
	+2F_{4,\pounds_{\bm{u}}\psi},
    \notag \\
    c_{10}
	&=
	2G_{4,\langle 22\rangle}
	-2F_{4,Z}\pounds_{\bm{u}}\psi,
    \qquad 
    c_{11}
	=
	-\frac{4G_{4,\langle 12\rangle}}{N}
	-2G_{4,\langle 22\rangle}
	\pounds_{\bm{u}}\psi
	-F_{4,\pounds_{\bm{u}}\psi}
	\pounds_{\bm{u}}\psi,
    \qquad
    c_{13}
	=
	-2G_{4,\langle 22\rangle},
    \notag \\   
    c_{14}
	&=
	-2F_{4,Z},
    \qquad 
    c_{15}=G_4,
    \qquad 
    d_{5}
	=
	2G_{4,\langle 22\rangle}
	+2F_{4,Z}\pounds_{\bm{u}}\psi.
\end{align}

Equations~\eqref{eq:Bi-Galileon in unitary}-\eqref{eq:L3} thus show
explicitly that, up to boundary terms, the bi-Galileon theory belongs to the
broad class \eqref{eq:S0}.

\section{$3+1$ decomposition of the interacting hypersurface approach}
\label{app:dec}

For a single scalar field $\phi$ with a timelike gradient, its level surfaces
define a preferred foliation, and one may use the temporal coordinate freedom
to impose the unitary gauge $\phi=t$. With two independent scalar fields
$\phi$ and $\psi$, a single temporal coordinate choice cannot in general make
both fields homogeneous. The construction in Sec.~\ref{sec:scgsf} therefore
chooses $\phi$ to define the foliation and imposes $\phi=t$, while $\psi$
remains explicit. A generally covariant bi-scalar-tensor theory can then be
written in spatially covariant form by using
Eqs.~\eqref{eq:dec_nbl1_psi} and \eqref{eq:dec_nbl2_psi}, together with the
Gauss-Codazzi relations.

By contrast, Ref.~\cite{Yu:2024sed} constructed bi-scalar-tensor theories
through an interacting-hypersurface approach, which extends single-field SCG
by introducing two independent foliation structures. In this appendix, we
show how the geometric quantities of that construction can be decomposed with
respect to a single foliation and expressed in the spatially covariant form
used in Eq.~\eqref{eq:S0}.

In the construction of Ref.~\cite{Yu:2024sed}, each scalar field defines
a foliation of hypersurfaces, denoted by $\Sigma_{\phi}$ and $\Sigma_{\psi}$,
respectively, on which the corresponding scalar field is uniform. Both
foliations are accompanied by their own geometric quantities. The
relevant geometric quantities are summarized in Table \ref{table: Basic geometric quantities}.
These quantities will serve as the basic building blocks for constructing
the two-field scalar-tensor theory.
\begin{table}[h]
	\centering 
	\begin{centering}
		\begin{tabular}{|c|c|c|}
			\hline 
			Geometric quantities  & w.r.t. $\Sigma_{\phi}$  & w.r.t. $\Sigma_{\psi}$\tabularnewline
			\hline 
			normal vectors  & $u_{a}\coloneqq-N\nabla_{a}\phi$  & $v_{a}\coloneqq-M\nabla_{a}\psi$\tabularnewline
			\hline 
			induced metrics  & $h_{ab}\coloneqq g_{ab}+u_{a}u_{b}$  & $l_{ab}\coloneqq g_{ab}+v_{a}v_{b}$\tabularnewline
			\hline 
			normalization factors  & $N\coloneqq1/\sqrt{2X}$  & $M\coloneqq1/\sqrt{2Y}$\tabularnewline
			\hline 
			accelerations  & $a_{a}\coloneqq\mathrm{D}_{a}\ln N$  & $b_{a}\coloneqq\tilde{\mathrm{D}}_{a}\ln M$\tabularnewline
			\hline 
			extrinsic curvatures  & $K_{ab}\coloneqq\frac{1}{2}\pounds_{\boldsymbol{u}}h_{ab}$  & $L_{ab}\coloneqq\frac{1}{2}\pounds_{\boldsymbol{v}}l_{ab}$\tabularnewline
			\hline 
			Ricci tensors  & $^{3}\!R_{ab}\equiv{}^{3}\!R_{ab}\left(h\right)$  & $^{3}\!\tilde{R}_{ab}\equiv{}^{3}\!\tilde{R}_{ab}\left(l\right)$\tabularnewline
			\hline 
		\end{tabular}
		\par\end{centering}
	\caption{Geometric quantities of two foliations of spacelike hypersurfaces.}
	\label{table: Basic geometric quantities} 
\end{table}
The action is constructed by coupling the two sets of geometric quantities.
An action built from these quantities can be written as
\begin{equation}
	S=\int\mathrm{d}^{4}x\sqrt{-g}\mathcal{L}\left(\phi,N,u_{a},h_{ab},{}^{3}\!R_{ab},\mathrm{D}_{a},\pounds_{\boldsymbol{u}};\psi,M,v_{a},l_{ab},{}^{3}\!\tilde{R}_{ab},\tilde{\mathrm{D}}_{a},\pounds_{\boldsymbol{v}}\right).
\end{equation}
Ref.~\cite{Yu:2024sed} imposed the restriction that there be no explicit
mixing of derivative operators and that temporal (Lie) derivatives enter only
at first order through the induced metrics. The resulting action takes the form
\begin{equation}
	S=\int\mathrm{d}^{4}x\sqrt{-g}\mathcal{L}\left(\phi,N,u_{a},h_{ab},K_{ab},{}^{3}\!R_{ab},\mathrm{D}_{a};\psi,M,v_{a},l_{ab},L_{ab},{}^{3}\!\tilde{R}_{ab},\tilde{\mathrm{D}}_{a}\right).\label{eq:bi-SCG}
\end{equation}

For our purpose, we decompose the geometric quantities of $\Sigma_{\psi}$
with respect to $\Sigma_{\phi}$.

The induced metric associated with $\phi$ is
$h_{ab}=g_{ab}+u_a u_b$. We decompose $v_a$ into components normal and
tangential to $\Sigma_\phi$ as
\begin{equation}
	v_{a}=h^{a^{\prime}}_{a}v_{a^{\prime}}-u^{a^{\prime}}u_{a}v_{a^{\prime}}=-\alpha u_{a}+\beta_{a},
\end{equation}
where $\alpha\equiv u^{a'}v_{a'}$ and
$\beta_a\equiv h^{a'}_a v_{a'}$. The quantity $\alpha$ is not independent.
Since $v_a v^a=-1$, and assuming that both $u^a$ and $v^a$ are
future-directed, we have
\begin{equation}
	\alpha=-\sqrt{1+\beta^{i}\beta_{i}}.
\end{equation}
In coordinates adapted to $\Sigma_\phi$, the components of $v_a$ and $v^a$
are therefore
\begin{align}
	v_{a} & =\left(\alpha N+N^{i}\beta_{i},\beta_{i}\right),\\
	v^{a} & =\left(-\frac{\alpha}{N},\alpha\frac{N^{i}}{N}+\beta^{i}\right).
\end{align}

The induced metric associated with $\psi$ decomposes as
\begin{equation}
	\begin{aligned}l_{ab} & =\left(h^{a^{\prime}}_{a}-u^{a^{\prime}}u_{a}\right)\left(h^{b^{\prime}}_{b}-u^{b^{\prime}}u_{b}\right)l_{a^{\prime}b^{\prime}}\\
		& =l_{\hat{a}\hat{b}}-u_{a}l_{\bm{u}\hat{b}}-u_{b}l_{\hat{a}\bm{u}}+u_{a}u_{b}l_{\bm{u}\bm{u}},
	\end{aligned}
\end{equation}
where 
\begin{align}
	l_{\hat{a}\hat{b}} & =h_{ab}+\beta_{a}\beta_{b},\\
	l_{\bm{u}\hat{b}} & =\alpha\beta_{b},\\
	l_{\bm{u}\bm{u}} & =-1+\alpha^{2}=\beta^{i}\beta_{i}.
\end{align}
Here and below, we follow the convention of
Refs.~\cite{Deruelle:2009zk,Deruelle:2012xv}. An index replaced by $\bm u$
denotes contraction with $u^a$, whereas a hatted index denotes projection by
$h_a{}^b$.

To decompose the extrinsic curvature $L_{ab}$, it is convenient first to
decompose $\nabla_a v_b$:
\begin{align}
	\nabla_{a}v_{b} & =\left(h^{a^{\prime}}_{a}-u^{a^{\prime}}u_{a}\right)\left(h^{b^{\prime}}_{b}-u^{b^{\prime}}u_{b}\right)\nabla_{a^{\prime}}v_{b^{\prime}}\nonumber \\
	& =\nabla_{\hat{a}}v_{\hat{b}}-u_{b}\nabla_{\hat{a}}v_{\bm{u}}-u_{a}\nabla_{\bm{u}}v_{\hat{b}}+u_{a}u_{b}\nabla_{\bm{u}}v_{\bm{u}},
\end{align}
where 
\begin{align}
	\nabla_{\hat{a}}v_{\hat{b}} & =-\alpha K_{ab}+\mathrm{D}_{a}\beta_{b},\\
	\nabla_{\hat{a}}v_{\bm{u}} & =\mathrm{D}_{a}\alpha-\beta^{b}K_{ab},\\
	\nabla_{\bm{u}}v_{\hat{b}} & =-\alpha a_{b}+\pounds_{\bm{u}}\beta_{b}-\beta^{a}K_{ab},\\
	\nabla_{\bm{u}}v_{\bm{u}} & =\pounds_{\bm{u}}\alpha-\beta^{a}a_{a}.
\end{align}

The extrinsic curvature $L_{ab}$ then decomposes as
\begin{align}
	L_{ab} & =\left(h^{a^{\prime}}_{a}-u^{a^{\prime}}u_{a}\right)\left(h^{b^{\prime}}_{b}-u^{b^{\prime}}u_{b}\right)L_{a^{\prime}b^{\prime}}\nonumber \\
	& =\left(h^{a^{\prime}}_{a}-u^{a^{\prime}}u_{a}\right)\left(h^{b^{\prime}}_{b}-u^{b^{\prime}}u_{b}\right)l^{c}_{b^{\prime}}l^{d}_{a^{\prime}}\nabla_{(c}v_{d)}\nonumber\\
	& =L_{\hat{a}\hat{b}}-2u_{(b}L_{\hat{a})\bm{u}}+u_{a}u_{b}L_{\bm{u}\bm{u}},
\end{align}
where
\begin{align}
	L_{\hat{a}\hat{b}} & =h^{d}_{a}h^{c}_{b}\nabla_{(c}v_{d)}+h^{d}_{(a}\beta_{b)}v^{c}\nabla_{c}v_{d}\nonumber \\
	& =-\alpha K_{ab}+\mathrm{D}_{(a}\beta_{b)}+\alpha^{2}a_{(a}\beta_{b)}-\alpha\beta_{(b}\pounds_{\bm{u}}\beta_{a)}+\beta_{(b|}\beta^{c}\mathrm{D}_{c}\beta_{|a)},
\\
	L_{\hat{a}\bm{u}}&=  h^{c}_{a}u^{b}\nabla_{b}v_{c}+\alpha h^{c}_{a}v^{b}\nabla_{b}v_{c}\nonumber \\
	&=  -\beta^{b}K_{ba}+\frac{1}{2}\left(\alpha\beta^{2}a_{a}-\beta^{2}\pounds_{\bm{u}}\beta_{a}+\alpha\beta^{b}\mathrm{D}_{b}\beta_{a}\right)\nonumber \\
	&\quad +\frac{1}{2}\left(-\beta_{a}\beta^{c}\pounds_{\bm{u}}\beta_{c}+\alpha\beta_{a}\beta^{b}a_{b}+\frac{1}{\alpha}\beta^{b}\mathrm{D}_{a}\beta_{b}+\frac{1}{\alpha}\beta_{a}\beta^{b}\beta^{c}\mathrm{D}_{b}\beta_{c}\right),
\end{align}
and
\begin{align}
	L_{\bm{u}\bm{u}} & =u^{a}u^{b}\nabla_{a}v_{b}+\alpha u^{b}v^{a}\nabla_{a}v_{b}\nonumber \\
	& =-\beta^{2}\pounds_{\bm{u}}\alpha+\beta^{2}\beta^{a}a_{a}+\alpha\beta^{a}\mathrm{D}_{a}\alpha-\alpha\beta^{a}\beta^{b}K_{ab}\nonumber \\
	& =\beta^{2}\beta^{a}a_{a}+\beta^{a}\beta^{b}\mathrm{D}_{a}\beta_{b}-\frac{\beta^{2}}{\alpha}\beta^{b}\pounds_{\bm{u}}\beta_{b}-\frac{1}{\alpha}\beta^{a}\beta^{b}K_{ab}.
\end{align}
In obtaining the final forms of $L_{\hat a\bm u}$ and $L_{\bm u\bm u}$,
we used
\begin{align}
	\mathrm{D}_{a}\alpha & =\frac{1}{\alpha}\beta^{b}\mathrm{D}_{a}\beta_{b},\\
	\pounds_{\bm{u}}\alpha & =\frac{1}{\alpha}\beta^{b}\pounds_{\bm{u}}\beta_{b}-\frac{1}{\alpha}\beta^{a}\beta^{b}K_{ab}.
\end{align}

Similarly, the acceleration $b_a$ decomposes as
\begin{equation}
	b_{a}=-u_{a}u^{a^{\prime}}b_{a^{\prime}}+h^{a^{\prime}}_{a}b_{a^{\prime}}=-u_{a}b_{\bm{u}}+b_{\hat{a}},
\end{equation}
where 
\begin{align}
	b_{\bm{u}} & =-\beta^{a}\pounds_{\bm{u}}\beta_{a}+\alpha a^{a}\beta_{a}+\frac{1}{\alpha}\beta^{a}\beta^{b}\mathrm{D}_{a}\beta_{b},\\
	b_{\hat{a}} & =-\alpha\pounds_{\bm{u}}\beta_{a}+\alpha^{2}a_{a}+\beta^{b}\mathrm{D}_{b}\beta_{a}.
\end{align}
Thus all the basic geometric quantities on $\Sigma_\psi$ can be decomposed
into components normal and tangential to $\Sigma_\phi$.

As pointed out in Ref.~\cite{Yu:2024sed}, the interacting-hypersurface
approach involves an important subtlety.
Since the two sets of geometric quantities are defined with respect
to two foliations specified by different timelike normal vectors $u^{a}$
and $v^{a}$, the spatial tensors on one foliation are no longer purely
spatial with respect to the other. The quantities on $\Sigma_\psi$ can be
decomposed into components normal and tangential to $\Sigma_\phi$ and
expressed in terms of $K_{ab}$ and the spatial vector
$\beta_a\equiv h^b_a v_b$ on $\Sigma_\phi$. Schematically,
\begin{equation}
	f\left(v_{a},l_{ab},L_{ab},{}^{3}\!\tilde{R}_{ab};\tilde{\mathrm{D}}_{a}\right)=g\left(\beta_{a},\pounds_{\bm{u}}\beta_{a},K_{ab},{}^{3}\!R_{ab};\mathrm{D}_{a}\right).
\end{equation}
Thus tensors that are spatial with respect to $\Sigma_\psi$ generally have
components normal to $\Sigma_\phi$.

As an illustration, we focus on the interaction $K_{ab}L^{ab}$ and consider
the model
\begin{equation}
	\mathcal{L}=c_{1}K_{ab}K^{ab}+c_{2}K^{2}+c_{3}K_{ab}L^{ab}+c_{4}{}^{3}\!R,\label{eq:model}
\end{equation}
where the coefficients $c_i$ are general functions of $N$, $M$, and
$\alpha\equiv u_a v^a$. Using the preceding results, the decomposition of
$K_{ab}L^{ab}$ with respect to $\Sigma_\phi$ is
\begin{equation}
	K_{ab}L^{ab}=g^{ac}g^{bd}K_{ab}L_{cd}=K^{ij}\left(-\alpha K_{ij}+\mathrm{D}_{j}\beta_{i}+\alpha^{2}a_{i}\beta_{j}-\alpha\beta_{j}\pounds_{\bm{u}}\beta_{i}+\beta_{j}\beta^{k}\mathrm{D}_{k}\beta_{i}\right),
\end{equation}
where $\beta_a=h^b_a v_b$ is the projection of $v_a$ onto $\Sigma_\phi$.
Using the definition of $v_a$, we find
\begin{equation}
	\beta_{i}=-\frac{1}{\sqrt{2Y}}\mathrm{D}_{i}\psi,\quad\text{with}\quad Y=-\frac{1}{2}(\partial_{a}\psi)^{2}\equiv\frac{1}{2}\left[\left(\pounds_{\bm{u}}\psi\right)^{2}-\mathrm{D}_{i}\psi\mathrm{D}^{i}\psi\right],\label{eq:beta_i-1}
\end{equation}
where $i=1,2,3$ labels the spatial coordinates $x^i$ adapted to
$\Sigma_\phi$. The second Lie derivative $\pounds^2_{\bm u}\psi$ enters
through $\pounds_{\bm u}\beta_i$, which is
\begin{align}
	\pounds_{\bm{u}}\beta_{i} & =-\frac{1}{\sqrt{2Y}}\pounds_{\bm{u}}\left(\mathrm{D}_{i}\psi\right)+\frac{1}{2\sqrt{2}}Y^{-\frac{3}{2}}\pounds_{\bm{u}}Y\mathrm{D}_{i}\psi\nonumber \\
	& =-\frac{1}{\sqrt{2Y}}\left(\mathrm{D}_{i}\left(\pounds_{\bm{u}}\psi\right)+\frac{\mathrm{D}_{i}N}{N}\pounds_{\bm{u}}\psi\right)
    \notag \\ & \quad 
    +\frac{1}{2\sqrt{2}}Y^{-\frac{3}{2}}\left[\pounds_{\bm{u}}\psi\pounds^{2}_{\bm{u}}\psi-\mathrm{D}^{j}\psi\left(\mathrm{D}_{j}\left(\pounds_{\bm{u}}\psi\right)+\frac{\mathrm{D}_{j}N}{N}\pounds_{\bm{u}}\psi\right)+K^{jk}\mathrm{D}_{j}\psi\mathrm{D}_{k}\psi\right]\mathrm{D}_{i}\psi.\label{eq:ln beta_i-1}
\end{align}
In deriving Eq.~\eqref{eq:ln beta_i-1}, we used
\begin{align}
\pounds_{\bm{u}}\left(\mathrm{D}_{i}\psi\right)&=\mathrm{D}_{i}\left(\pounds_{\bm{u}}\psi\right)+\frac{\mathrm{D}_{i}N}{N}\pounds_{\bm{u}}\psi,\label{eq:ln Di psi}
\\
\pounds_{\bm{u}}\left(\mathrm{D}^{i}\psi\right)&=\mathrm{D}^{i}\left(\pounds_{\bm{u}}\psi\right)+\frac{\mathrm{D}^{i}N}{N}\pounds_{\bm{u}}\psi-2K^{ij}\mathrm{D}_{j}\psi,
\end{align}
and
\begin{align}
	\pounds_{\bm{u}}Y & =\pounds_{\bm{u}}\psi\pounds^{2}_{\bm{u}}\psi-\frac{1}{2}\pounds_{\bm{u}}\left(h^{ab}\mathrm{D}_{a}\psi\mathrm{D}_{b}\psi\right)\nonumber \\
	& =\pounds_{\bm{u}}\psi\pounds^{2}_{\bm{u}}\psi-\mathrm{D}^{i}\psi\left(\mathrm{D}_{i}\left(\pounds_{\bm{u}}\psi\right)+\frac{\mathrm{D}_{i}N}{N}\pounds_{\bm{u}}\psi\right)+K^{ij}\mathrm{D}_{i}\psi\mathrm{D}_{j}\psi.\label{eq:ln Y}
\end{align}

Equation~\eqref{eq:ln beta_i-1} shows explicitly that a quantity containing
only first Lie derivatives in the $\psi$-foliation description can generate
$\pounds^2_{\bm u}\psi$ when rewritten relative to $\Sigma_\phi$. The
appearance of this higher time derivative does not by itself establish an
additional degree of freedom or a ghost: the answer depends on the full
nonlinear constraint structure. Nevertheless, a single temporal coordinate
choice can generically impose unitary gauge on only one of two independent
scalars, so the higher-time-derivative dependence of both fields cannot in
general be removed simultaneously by a gauge choice. The health of the
general interacting-hypersurface action must therefore be assessed by an
appropriate degeneracy and constraint analysis. In the single-foliation
description adopted in this paper, this observation motivates the broad
action \eqref{eq:S0} and the Hamiltonian analysis of Secs.~\ref{sec:ham} and
\ref{sec:dccond}.

\providecommand{\href}[2]{#2}\begingroup\raggedright\endgroup

\end{document}